\documentclass[a4paper,12pt]{article}
\usepackage{graphicx}
\usepackage{dcolumn}
\usepackage{bm}
\usepackage{amssymb}
\usepackage{amsmath}
\usepackage{hyperref}
\usepackage{authblk}
\usepackage{cite}
\usepackage{float}
\usepackage{caption}

\begin{document} 

\title{Correlation Integral  vs.~second order Factorial Moments \\
and an efficient computational technique }
\author[1]{F.~K.~Diakonos\footnote{email: fdiakono@phys.uoa.gr}}
\author[2]{A.~S.~Kapoyannis\footnote{email: akapog@phys.uoa.gr}}
\affil[1,2]{Nuclear and Particle Physics Section, Faculty of Physics, 
University of Athens, GR-15784 Greece}
\date{}

\maketitle

\begin{abstract}
We develop a mapping between the factorial moments of the second order $F_2$ and the correlation
integral $C$. We formulate a fast computation technique for the evaluation of both, which is 
more efficient, compared to conventional methods, for data containing number of pairs per event
which is lower than the estimation points. We find the effectiveness of the 
technique to be more prominent as the dimension of the embedding space increases.
We are able to analyse large amount of data in short computation time and access very low scales
in $C$ or extremely high partitions in $F_2$. The technique is an indispensable tool for detecting 
a very weak signal hidden in strong noise.
\end{abstract}



\section{Introduction}\label{sec:intro}

Factorial moment analysis has been introduced in \cite{mom1,mom2} as a very promising tool to study correlation phenomena in particle 
physics. In particular, it has been argued in several works 
\cite{inter1,inter2,inter3,inter4,inter5,inter6,inter7,inter8,inter9,inter10,inter11,inter12,inter13,inter14,
inter15,inter16} 
that long range correlations related to critical behaviour can be detected through the occurrence of 
power-law behaviour of the factorial moments as a function of the scale. In particle physics, this behaviour, called intermittency,
is expected to occur at 
very small momentum differences as a manifestation of the phenomenon of critical opalescence \cite{Crit-Opal}. However, the 
calculation of the factorial moments at very small scales requires a huge computational effort which prevents its implementation in 
standard experimental analysis, restricting the related calculations to intermediate scales, at which the phenomenon may be lost. In 
the present work we provide an alternative way to overcome this difficulty proposing at the same time a novel protocol for 
fluctuation analysis. Our scheme can be applied in general point sets, not necessarily related to particle physics. Therefore, the 
discussion in the following tries to adopt this very general framework of point set analysis. Nevertheless, the procedure concerning 
the application of the method to particle physics tasks is made sufficiently transparent.

\begin{figure*}[h]
\centering
\includegraphics[scale=0.5,trim=3.6cm 8.cm 3.6cm 1.cm,angle=0]{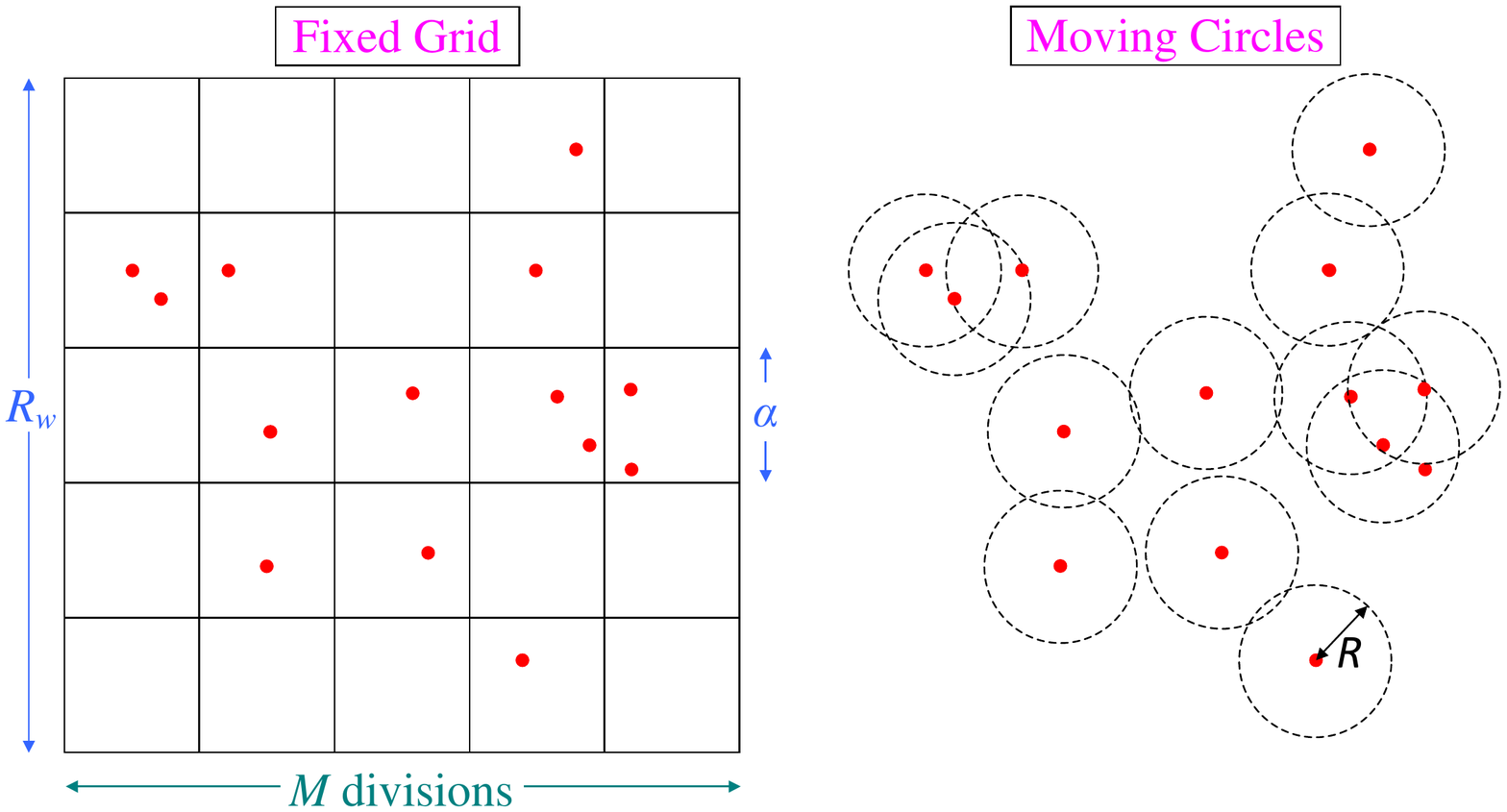}\\
\vspace{-0.4cm}
\hspace{2.8cm}(a)\hspace{6.2cm}(b)\hspace{7.5cm}
\vspace{-0.2cm}
\vspace{-0cm}
\caption{\label{fig:Grid-Circles} {\small The calculation procedure of factorial moments of
second order, $F_2$, in (a) and correlation integral, $C$, in (b).}}

\end{figure*}

Conventionally, the factorial moments of the second order $F_2$ can be evaluated as
\[
F_2 (M)=\frac{{\frac{1}{{{N_{e}}}}\sum\limits_{{N_{e}}} {\frac{1}{{{M^d}}}\sum\limits_{c} {N(N - 1)} } }}{\left( {\frac{1}{N_e}\sum\limits_{{N_{e}}} {\frac{1}{{{M^d}}}\sum\limits_{c} N } } \right)^2}=
\frac{{\frac{1}{{{M^d}}}\sum\limits_{c} {{{\left\langle {N(N - 1)} \right\rangle }_{e}}} }}{{{{\left( {\frac{1}{{{M^d}}}\sum\limits_{c} {{{\left\langle N \right\rangle }_{e}}} } \right)}^2}}} 
\]
\begin{equation}\label{eq:F2}
= M^d \frac{\sum\limits_c \left\langle {N(N - 1)} \right\rangle _e }
{\left( {\sum\limits_c \left\langle N \right\rangle _e } \right)^2} =
{M^d}\frac{{\sum\limits_{c} {{{\left\langle {N(N - 1)} \right\rangle }_{e}}} }}{{\left\langle {{N_{m}}} \right\rangle _{e}^2}}\;,
\end{equation}
where $N_e$ is the number of events in a data set and $N_m$ the multiplicities per event.
To calculate $F_2$ we form a fixed subspace in the $d$-dimensional embedding space, defined by sides
$R_w$, having total volume $R_w^d$, which encloses the whole set of points we want to analyse. 
For two dimensions this a fixed window of initial size $R_w$, as shown in 
Fig.~\ref{fig:Grid-Circles}(a).
We then divide each side in $M$ equal segments and form a grid of cells $c$. We count the points $N$
that fall within one cell and sum over all the cells.
The factorial moments of the 2nd order for a fractal data set of dimension $d_F$ behave as function
of $M^d$ as
\begin{equation}\label{eq:F2_propto}
{F_2}(M^d) \propto M^{d -d_F} = 
\left( M^d \right)^{ 1-\frac{d_F}{d} }\;,
\end{equation}
revealing, thus, by their slope $1-d_F/d$ in a logarithmic diagram, the fractal dimension $d_F$.

The other method to address correlations is by the correlation integral, $C$, \cite{CI}. This quantity
reveals how a system is structured at different scales $R$, which represent differences between values
of any physical quantity that we want to study\footnote{Thus, the units of $R$ are the units of the
relevant physical quantity and will be left arbitrary throughout this paper} and it is evaluated
to be
\begin{equation}\label{eq:CI}
C(R) = \frac{2}{\left\langle{N_m\left(N_{m-1}\right)}\right\rangle_e}
{\left \langle \sum_{i=1}^{N_m} \sum_{j=1}^{i-1}      
\theta (|\vec{x}_i - \vec{x}_j| \le R)  \right\rangle _e}\;.
\end{equation}
Here, for a specific scale $R$ we form freely moving disks\footnote{With the term disk 
throughout this paper we mean $d$-dimensional objects which contain all the points of the space with
distances, $s_D$, from a certain point which is the center, obeying $s_D \le R$.} with radius $R$ which is measured from 
the centre of the disk. The disks are line segments, circles or spheres in an embedding space of 
one, two or three dimensions, respectively. We position the centre of the disk at each point $i$ of 
our data with coordinates $\vec{x}_i$ and count the number of points enclosed by the disk.
This is carried out by evaluating the distance $s_D$ of all the rest points $j$ with coordinates 
$\vec{x}_j$ from the point $i$, which in a $d$-dimensional embedding space, is 
$s_D \equiv |\vec{x}_i - \vec{x}_j|=
\left[ \sum\limits_{k=1}^{d} \left( x_{ik}-x_{jk} \right)^2 \right]^{1/2}$. The distance $s_D$
is compared with the scale $R$. 
The procedure for two dimensions is shown in Fig.~\ref{fig:Grid-Circles}(b).
The correlation integral for a fractal data set of dimension $d_F$ behaves as function
of $R$ as 
\begin{equation}\label{eq:CI_propto}
C(R) \propto R^{d_F}\;,
\end{equation}
which has slope in a logarithmic diagram equal to the fractal dimension $d_F$.

Thus, the quantities $F_2$ and $C$ are interrelated since they determine the fractal dimension of a point set.
As we will show in the present work, it is possible to obtain a direct relation between $F_2$ and $C$. This 
relation turns out to be very useful whenever the correlation analysis is performed in an ensemble of point 
sets each containing a small number of points. Such a situation necessarily occurs in applications in particle
physics when the correlations between particles of a specific species are calculated through averaging over
a large number of events. Then, it often appears the case that the number of particles of interest 
per event -and consequently the set of their coordinates which form the considered point set- is very small. 
As a typical example we refer the reader to the correlation analysis of proton transverse momenta in ion
collisions \cite{NA49_protons}. Then, the calculation of $F_2$ becomes prohibitive as the size of cells in the embedding
space becomes very small. Here we will show how we can profit from the derived relation between $F_2$ and $C$ 
in order to develop a very efficient computational tool allowing to perform calculations in this prohibited
region which is crucial for the detection of critical correlations
\cite{Antoniou_PRC_2016,Antoniou_jopg_2019}.

The structure of this paper is as follows. In section \ref{sec:map} we establish a relation
between $C$ and $F_2$ which allows to pass our calculations from one to the other. We,
also, stress the advantages by working with the correlation integral.
Further issues, related to this correspondence in cases of data of varying multiplicities, are
discussed in \ref{sec:dif_mul}. 
In section \ref{sec:ring} we develop a fast computational algorithm for calculating the correlation
integral which is, also, passed to $F_2$. We probe its effectiveness with respect to conventional
methods of calculation.
In section \ref{sec:app} we apply the $C-F_2$ correspondence and our calculation technique to
analyse several
data sets, which reside at embedding spaces with one to three dimensions. We, also, note the limitations which
a finite data set exhibits.
In section \ref{sec:micro} we study a situation where a data set contains the needed information 
with a large percentage of unwanted ``noise'' and show how our method can reveal this information.
We analyse a simulated data set with these attributes and show that conventional techniques would
be inadequate.
In section \ref{sec:conclu} we summarise our conclusions.
Our paper is accompanied by the source computing code, as supplemental file, which can be used for the relevant calculations.

\section{Mapping between $F_2$ and {\it C}}\label{sec:map}

The factorial moments of the second order, $F_2$ (\ref{eq:F2}) and the correlation integral, $C$ 
(\ref{eq:CI}) show an underlying
similarity. 
Both measure the number of pairs in a set of points at some scale. Indeed, to calculate 
$F_2$, one has to count the number
of points $N$ that fall inside the cells of a certain partition $M$. Then, in eq.~(\ref{eq:F2}) the 
number $N(N-1)$ is recorded, which is twice the number of pairs.
Also, in the calculation of $C$, one has to count the number $N$ of points that fall within a
disk of radius $R$ centred at a specific point. 
This is the number of pairs that can be formed with this specific point. The procedure 
is repeated for all the points, so
that in the end we have all the pairs that can fit within disks of radius $R$. This similarity allows 
us to map between $F_2$ and $C$. 

To find the average total number of pairs, $\left\langle N_p(M)\right\rangle_e$, used in $F_2$, which 
corresponds to a partition $M$, 
 we have to add for
all cells of this partition the average quantity
 $\left\langle N(N - 1)\right\rangle_e$ of each
cell, which appears in eq.~(\ref{eq:F2}). Solving the equation for this
quantity and having in mind that the pairs in $F_2$ are counted twice, we get
\begin{equation}\label{eq:Np_F2}
\left\langle N_p(M)\right\rangle_e=\frac{1}{2}
\sum\limits_{c} {{{\left\langle {N(N - 1)} \right\rangle }_{e}}} =
\frac{1}{2} 
 \frac{{\left\langle {{N_{m}}} \right\rangle _{e}^2}}{{{M^d}}}{F_2}(M)\;.
\end{equation}
The pairs which correspond to a scale $R$, are recorded in $C$ through 
the theta function appearing in eq.~(\ref{eq:CI}). This function counts
one for each point $j$ falling within the disk of radius $R$ centred at
point $i$ and zero otherwise. Thus, the average number of pairs, $\left\langle N_p(R)\right\rangle_e$, counted in 
$C$ and corresponding to the scale $R$, 
 are taken from eq.~(\ref{eq:CI}) to be
\begin{equation}\label{eq:Np_C}
\left\langle N_p(R)\right\rangle_e=
{\left \langle \sum_{i=1}^{N_m} \sum_{j=1}^{i-1}      
\theta (|\vec{x}_i - \vec{x}_j| \le R)  \right\rangle _{e}}=
\frac{{{{\left\langle {{N_{m}}\left( {{N_{m}} - 1} \right)} 
\right\rangle }_{e}}}}{2} C(R)\;.
\end{equation}
We, now, demand
the pairs which correspond to $C$ at the scale $R$ to be equal to the pairs
which correspond to $F_2$ at the partition $M$.
Then, with the help of eqs.~(\ref{eq:Np_F2}) and (\ref{eq:Np_C}),
we get
\[
\left\langle N_p(M)\right\rangle_e=
\left\langle N_p(R)\right\rangle_e \Rightarrow
\] 
\vspace{-0.5cm}
\begin{equation}\label{eq:eq_pairs}
\Rightarrow \frac{1}{2}
\frac{\left\langle N_m \right\rangle _e^2}{M^d} F_2(M)=
\frac{\left\langle N_m\left( N_m - 1 \right) \right\rangle _e}{2} C(R) \Rightarrow
C(R) =
\frac{\left\langle N_m \right\rangle _e^2}
{\left\langle N_m \left( N_m - 1 \right) \right\rangle _e}
\frac{1}{M^d} F_2(M)\;.
\end{equation}
If the number of multiplicities per event are constant eq.~(\ref{eq:eq_pairs}) is reduced to
\begin{equation}\label{eq:eq_pairs_red}
C(R) = \frac{N_m}{N_m - 1} \frac{1}{M^d} F_2(M)\;.
\end{equation}

We have to relate the scale $R$ to the partition $M$. 
Let our working space in the evaluation of $F_2$ have measure
 $V_w=R_w^d$, with $d$ being the embedding space dimension.
The measure of each cell in the partition $M$ is $V_{F_2,d}=a^d$.
Obviously $a=\frac{R_w}{M}$. Then it should hold that
\begin{equation}\label{eq:eq_R-M}
R \propto a = \frac{R_w}{M}\;.
\end{equation}
In the evaluation of $C$, the scale $R$ is a radius which defines distance around a certain point
in order to search for other points which are enclosed within the boundaries of the disk.
This measure is $V_{C,1}=2R$ in one dimension as line segment, 
$V_{C,2}=\pi R^2$ in two dimensions as circle and
$V_{C,3}=\frac{4}{3}\pi R^3$ in three dimensions as sphere.
Demanding equal measures for $F_2$ and $C$ we have that
\begin{equation}\label{eq:eq_space}
V_{F_2,d} = V_{C,d}\;.
\end{equation}
Substituting $V_{F_2,d}$ in eq.~(\ref{eq:eq_space}) we find the exact relation of the scale
$R$ to the partition $M$, for each space dimension $d$
\begin{equation}\label{eq:eq_spaces_compact}
R = \beta_d \frac{R_w}{M},\;\;\beta_1=\frac{1}{2},\;\beta_2=\frac{1}{\sqrt{\pi}},\;\beta_3=\sqrt[3]{\frac{3}{4\pi}}\;.
\end{equation}

Eq.~(\ref{eq:eq_pairs}) or (\ref{eq:eq_pairs_red}) with eqs.~(\ref{eq:eq_spaces_compact}) allow for the 
mapping between correlation integral and the second order factorial moment as
\begin{equation}\label{eq:C-F2}
C(R) =
a_m^{-1} \left(\frac{R}{\beta_d R_w}\right)^d F_2\left(\frac{\beta_d R_w}{R}\right)\;,
\end{equation}
\begin{equation}\label{eq:F2-C}
F_2(M) =
a_m M^d C\left(\frac{\beta_d R_w}{M}\right)\;,
\end{equation}
where $a_m$ is given by
\begin{equation}\label{eq:a_m}
a_m=\frac{\left\langle N_m \left( N_m - 1 \right) \right\rangle _e}
{\left\langle N_m \right\rangle _e^2}\;,\;\;\;
a_m=\frac{N_m - 1}{N_m},
\end{equation}
for varying or constant multiplicities $N_m$ per event, respectively.
In the Appendix we elaborate more specifically how to deal with situations where
our data contain events with different multiplicities $N_m$.

The above relation (\ref{eq:eq_spaces_compact}) 
between scales $R$ and partitions $M$ holds exactly at scales well below the scale 
which corresponds to the size of the ``box'' containing all the data so that
boundary effects are negligible. To understand why, let us consider
in $d=$2 dimensions a cell of size $a$ which can contain data points
and all around it exist other cells which, also, can contain 
data points. The corresponding disk is of radius $R=a/\sqrt{\pi}$
and is movable so that one point is placed at its center to count 
other points enclosed within. Two points may exist within the cell
which have distance $a\sqrt{2}$ and the cell will count them. The 
corresponding disk cannot count them, since their distance is greater
than the radius of the disk. However, if the disk, which is freely 
moving, is placed with its center at one point
within the cell it can find another point in one
of the adjacent cells
to form a pair within the disk. The outcome is 
equivalent, since the cell and the disk have equal surfaces, and thus
equal probability of finding points enclosed within
their domain.
The situation is altered, though, at low partitions.
Let us suppose that we prepare a set of points existing within a
window of side $R_w$. According to eq.~(\ref{eq:eq_spaces_compact}), 
the corresponding disk
has radius $R=R_w/\sqrt{\pi}$. A point placed at one corner of the
window, which represents our unique cell for $M=1$, can form a pair with every other point within the window.
But if we place the disk with its center close to one corner of the 
window it cannot count as partner of the pair a point with greater
distance than its radius. Now, the disk cannot find another point
outside the window to form a pair, since the window encloses all the existing data points.
As $M$ increases the boundary cells cover a decreasing
 percentage of the whole surface and so the boundary effects
diminish. 
To deal with this effect, when calculating $F_2$ through $C$, 
we assign to the $M=1$ partition not only the pairs  
enclosed within disks of radius $R=R_w/\sqrt{\pi}$, but, also, all
the pairs outside this disk. In this way the $M=1$ result of $F_2$ 
calculated from the grid and from the $C$ will always coincide.
For the few low $M$ partitions where the grid and $C$ calculations
may differ due to boundary effects, we can shift slightly the boundary
between the scales which correspond to adjacent $M$. According to
eq.~(\ref{eq:eq_spaces_compact}), an integer $M^\prime$ will correspond to an exact value of 
scale $R^\prime$ and an integer $M^{\prime\prime}=M^\prime-1$ will correspond 
to an exact value of scale $R^{\prime\prime}>R^\prime$. 
Normally, every scale $R$ obeying $R^{\prime\prime}>R>R^\prime$ should, also,
correspond to $M^\prime-1$. Instead, we can extend slightly the 
interval of scales corresponding to $M^\prime$, so that the $F_2$
grid calculations coincide, on the average, with the calculations through
$C$ at the low $M$ partitions\footnote{This is carried out through the
parameter $m_f$ introduced in eq.~(\ref{eq:M-m_f}) in 
the next chapter.}. As $M$ increases this extension will
have no effect, as the interval $R^{\prime\prime}-R^\prime$ shrinks.

However, apart from the similarity that enables the establishment of a correspondence between 
$F_2$ and $C$, 
crucial differences between the factorial moments and the correlation integral
{\it do exist}, which make apparent the advantages we have by working with $C$. These differences
are the following:

a) The grid, which is needed to calculate $F_2$ at partitions $M$, is fixed and not
directly related to the data. The results we get depend on the grid location relative to the data
and the size of the analysis space, controlled by $R_w$. 
In contrast, in order to calculate $C$, we form disks which locate 
themselves according to the data, since each disk has as its center, each time, a data point.
Consequently, in $C$ we always get a unique result for the same data set, whilst in $F_2$ our
result is grid depended.

b) Depending on the location of the grid we form to calculate $F_2$ at a specific partition 
$M$, a group of points may or may not be included within a cell. For this reason we may repeat the
calculation for the same $M$ with grids slightly shifted in different directions and then take as
result the average. This ``grid averaging'', on one hand, severely increases the time needed to
complete the calculation. On the other hand, the grid averaging does not ensures us that we have 
counted the maximum possible number of points per cell. During the averaging one grid may split a
group of points which could be fitted within the size of a cell of the specific partition, while 
another grid may split another group of points. Then, when we take the average, the result 
will be lower than the possible maximum value. This effect may be important when we are dealing
with events with low multiplicities.
On the contrary, with the moving disks of $C$ we never lose points at a specific scale $R$.
Our result is always the same (at the highest possible value), so we do not need to perform any
kind of averaging, saving, also, computation time.

c) The grid of $F_2$ entangles scales. For example, let us consider a partition $M$ with
cells of size $a$ in two dimensions. Then two points, located close to the diagonal corners of a cell
and separated by distance $s_D \simeq a\sqrt{2}$, are enclosed within the cell and counted for the estimation
of $F_2$. However, if two points exist in the data set, which have the same distance but their
connecting line is positioned paralleled to one grid axis, then they cannot be enclosed in
a cell of size $a$, no matter where we move the grid parallel to its axes.
Also, even when $s_D<a$ the two points may be separated in different cells. 
So a pair of points with a specific distance $s_D$ may or may not belong to a specific partition $M$.
Unlike $F_2$, $C$ provides a pure correspondence of the distance of a pair to the scales $R$.
There is a clear ``cut'' for the number of pairs $N(R)$ which belong to a specific scale $R$.
A pair with distance $s_D=R$ will always  be enclosed to measuring disks with radius $R' \ge R$ and, so,
it can be assigned unambiguously to all scales $>R$.

d) The grid calculation of $F_2$ introduces larger errors with the respect to $C$.
The source of these errors may be the possible splitting of pairs and the entanglement of scales.
Also, the odd or even partitions $M$ may introduce artificial fluctuations. This occurs, for example
if there is a large concentration of data points at some space location, e.g. at the center of the
grid. Then, if an odd partition encloses most of these points in a cell, then the even partition
will systematically split most of them, leading to considerably different results.
Counter to $F_2$, $C$ calculations, in general, provide smoother distributions, with lower errors.
This can be important at situations where the detection of a weak signal is needed.

e) The partitions $M$ which are estimation points for $F_2$ is a natural number. 
Thus, we estimate $F_2$ at discrete points. This does not matter, of course, at large $M$, but
at low partitions, close to unit, we have discontinuities.
In contrast $C$ can be calculated at any scale $R$, so we can have practically a continuous calculation
even for the large scales, which correspond to low $M$.

f) There is a difference which has no effect on the actual calculation, but it is related to
the physical understanding we may have on the system. The correlation integral $C$ reveals 
how the system behaves at different scales, which are direct physical quantities.
On the contrary, the results of $F_2$ are connected to the partitions $M$, which are
artificial quantities, giving us an indirect sense of the scales. 
The physical quantities are the sizes of the cells, with which,
qualitative, $M$ are inversely related. To reach at the exact quantitative relation between $M$ and
scales, we have to utilize the size of the whole space where we perform our analysis.

Finally, we note that the correlation integral, which is connected to distances between 
pair of points (doublets), has been related here to the factorial moments of the second order, which,
also, count pairs. Therefore, $C$ cannot be related to higher factorial moments,
$F_q$, $q=3,4,\ldots$, since they describe how the number of higher multiplets (triplets, quadruplets, etc.) change as scale (or partition) varies.

\section{A fast computational technique}\label{sec:ring}

In a conventional algorithm of calculation of 
$F_2$, one first has to form a specific grid of partition $M$. 
Then, for every event, the points in every cell have to be counted.
In the calculation of $C$, one first has to define a specific scale $R$. 
Then, for every event, disks of radius $R$ have to be formed
centred at each point and the other points which reside within the disk have to be counted.
In sort, a conventional algorithm first forms a grid with a certain partition or disks of 
certain scale and then counts the points that correspond to this partition or scale.

In the previous section we saw that the correlation integral, in contrast to the factorial moments, 
allows for a clear ``cut'' in the correspondence of pair of points to specific scales $R$.
This attribute allows for an improvement of the algorithm we shall use to calculate $C$. We can assign 
each pair to a 
specific scale $R$. In this way we know in advance all the disks where this specific pair belongs to. 
These are all the disks with radius $\geq R$. 
In the 2-dimensional case, we know that this pair fits exactly in a ``ring''{\footnote{With the term ring 
throughout this paper
we mean $d$-dimensional objects which contain all the points of the space with distances, $s_D$, from a
certain point which is the center, obeying $R-dR < s_D \le R$.} with radius
about $R$, as it is depicted in Fig.~\ref{fig:Ring+Circle}(a).
This picture of a 2-dimensional ring can be  
generalised in one dimension, where we have two line segments, each with length $dR$,
located symmetrically at distance $R$ from a point and in three dimensions, where we have a spherical
shell of thickness $dR$ and radius $R$. However, we shall call the technique we are developing with the name ring for every
dimension $d$.

\begin{figure*}[h]
\centering
\includegraphics[scale=0.4,trim=6.4cm 19.8cm 6.4cm 0.2cm,angle=0]{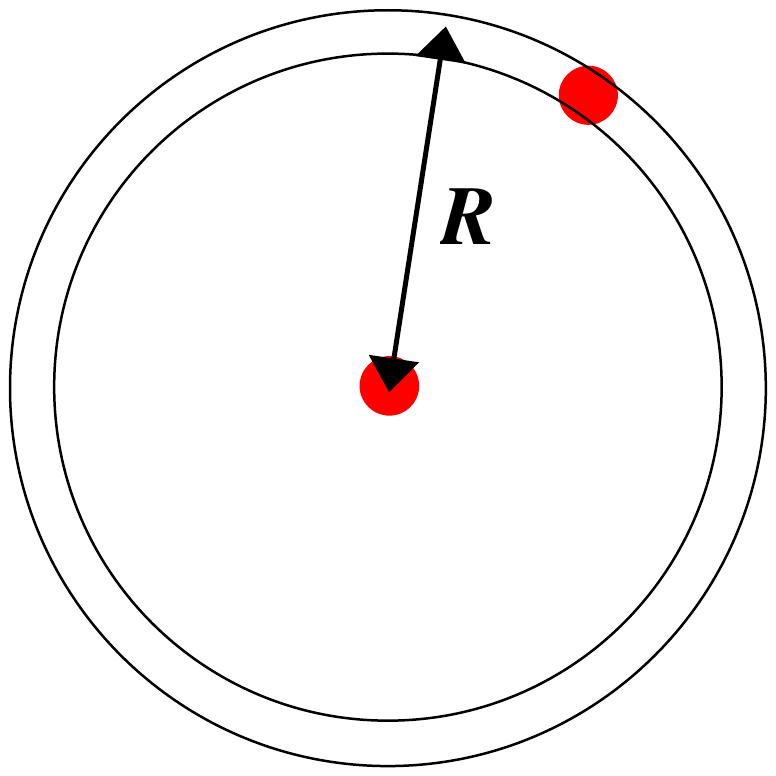}
\hspace{2cm}\includegraphics[scale=0.50,trim=3.3cm 12.8cm 3.3cm 0.cm,angle=0]{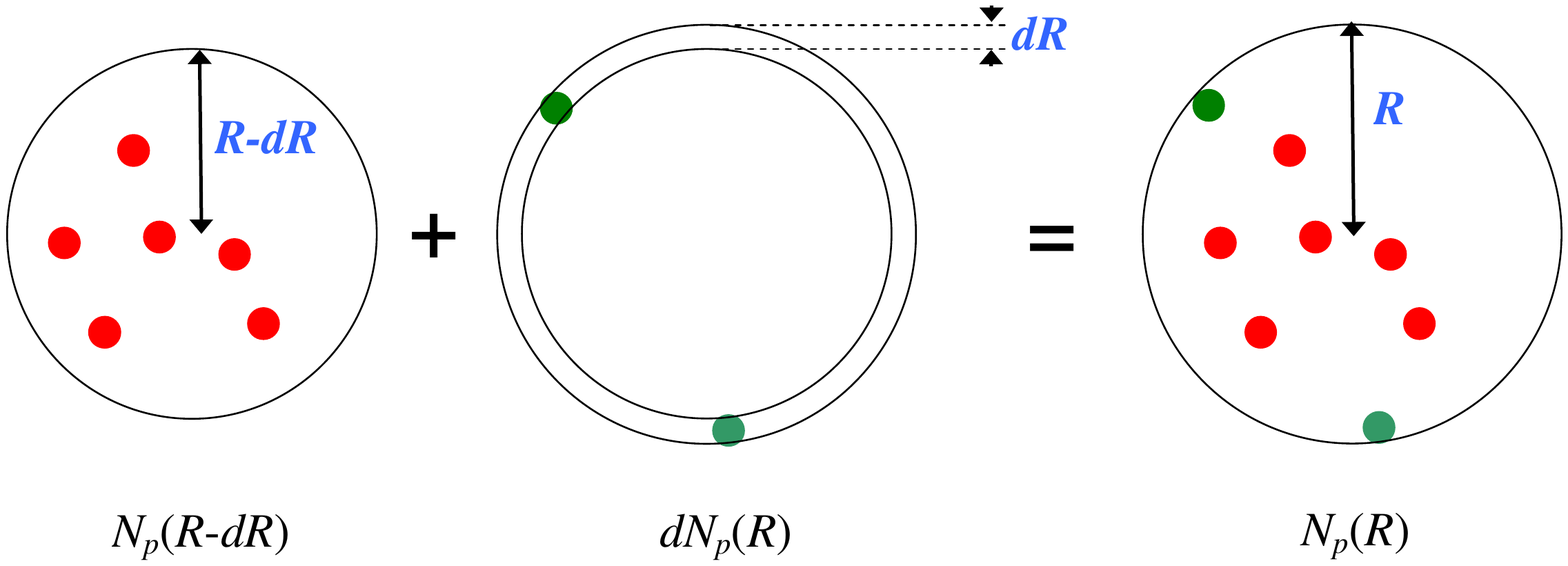}\\
\vspace{-0.4cm}
\hspace{0.5cm}(a)\hspace{4.5cm}(b)\hspace{11.5cm}
\vspace{-0.2cm}
\vspace{-0cm}
\caption{\label{fig:Ring+Circle} {\small (a) Two points at distance $R$ form a pair that fits
in a ring of radius $\sim${\it R} centred at one of the points.
(b) Here each pair of points is represented by one point, the one that is not positioned at the
centre of the ring or disk, so, the dots represent pairs. 
The ring or disks appearing in this figure represent all similar rings or disks
of equal size formed to analyse our data set, they are not positioned at a fixed location, 
but instead they have collected all the
points that fall within the domain they define. To count the pairs within the
disk $R$ we can count the pairs of disk $R-dR$ and add the pairs within the ring with
radii $R-dR$ and $R$.}}
\end{figure*}

So, in a data set we can form all possible pairs within the event, for all events,
calculate the distance between the two points of
the pair and then assign each pair to the
relevant ring. 
In the end of the process we have the number of pairs $dN_p(R)$ which correspond to this particular ring.
To find the number of pairs which correspond to a set of disks with radius $R$ we have to sum all the rings with
radius up to $R$. Equivalently, as it is described in Fig.~\ref{fig:Ring+Circle}(b), if we know the number of pairs
for the set of disks $N_p(R-dR)$, we simply have to add the pairs of the ring $dN_p(R)$ to find the number of pairs in the
set of disks $N_p(R)$.
The underlying physical meaning here is that the disks of radius $R$ in $C$ are connected to the cumulative probability function,
while the rings of thickness $dR$ are connected to the probability density.

In general, we can define as many rings as the number of our estimation points, $N_D$, at which we want 
to calculate $C$.
Thus, it becomes obvious the advantage we have through this technique.  With only one reading of data we have the value of
$C$ at all scales, whilst in the conventional way we had to read all the data for each scale for 
which we wanted to calculate $C$.
As a result in the new technique, the time needed to complete the calculation is little influenced by $N_D$.
The average number of pairs in a ring due to the events is
\begin{equation}\label{eq:dN(R)_e}
\left\langle dN_p(R)\right\rangle_e = \frac{1}{N_e} \sum_e dN_p(R),
\end{equation}
where the sum is formed with one reading to all $N_e \times \frac{1}{2}N_m \cdot(N_m-1)$ pairs of the data set.
Then the average number of pairs corresponding to scale $R$ is simply
\begin{equation}\label{eq:N(R)_e}
\left\langle N_p(R)\right\rangle_e = \sum_{r\leq R} \left\langle dN_p(r)\right\rangle_e  
=\left\langle N_p(R-dR)\right\rangle_e + \left\langle dN_p(R)\right\rangle_e\;.
\end{equation}
Then, through eq.~(\ref{eq:Np_C}), we can evaluate $C$ as 
\begin{equation}\label{eq:C_Np}
C(R)=\frac{2}{\left\langle N_m \left( N_m - 1 \right)
\right\rangle_e}
\left\langle N_p(R)\right\rangle_e\;.
\end{equation}

However, the situation becomes a little more complicated if we want to calculate the errors due to fluctuations imposed by the
different events. To accomplish this, we have to evaluate the error in the number of pairs within
the disk of radius $R$, due to different events. This error reads
\[
\delta \left\langle N_p(R)\right\rangle_e = 
\sqrt {\frac{1}{N_e-1} \left\langle \left[N_p(R)-\left\langle N_p(R)\right\rangle_e \right]^2\right\rangle_e} =
\]
\begin{equation}\label{eq:deltaN(R)}
=\sqrt {\frac{1}{N_e(N_e-1)} \left\{\sum_e \left[N_p(R)\right]^2 - \frac{1}{N_e}\left[ \sum_e N_p(R)\right]^2 \right\}}\;.
\end{equation}
The sum over the number of pairs in each ring per event, $N_p(R)$, for all events can be found by adding the results of all events and 
so it requires only one reading of all data.
But the sum of the squares of the number of pairs in each ring per event, $\left[N_p(R)\right]^2$, entails to
find for every event separately the number $N_p(R)$.
Consequently, after putting the $\frac{1}{2}N_m (N_m-1)$ pairs, available in an event, into rings, 
we have to read all the available $N_D$ 
rings to find out the result.
This process performs data reading $N_D \times N_e \times \frac{1}{2} N_m (N_m-1)$ times.
Despite this increase in processing time, the process remains advantageous, as we shall see later on.
The total error in $C$, using eq.~(\ref{eq:C_Np}), can be found to be
\begin{equation}\label{eq:deltaC}
\delta C(R)=C(R)
\sqrt{
\left[\frac{\delta \left\langle N_p(R)\right\rangle_e}
{\left\langle N_p(R)\right\rangle_e}\right]^2+
\left[\frac{\delta \left\langle N_m \left( N_m - 1 \right)\right\rangle_e}
{\left\langle N_m \left( N_m - 1 \right)\right\rangle_e}\right]^2
}\;,
\end{equation}
where we have included the case of varying $N_m$ per event. 

Through mapping between $F_2$ and $C$ we can use the above technique to calculate $F_2$, as well.
Thus, we can calculate $C$ and then, using eq.~(\ref{eq:eq_pairs}) or (\ref{eq:eq_pairs_red}), we can evaluate $F_2$.
More directly, we can follow similar steps with the
ones in the calculation of $C$ with the ring
technique.
The application of this technique in this case, 
again, involves finding the distances between all the
pairs in the data set with only one reading of
data. Through eqs.~(\ref{eq:eq_spaces_compact})
we can assign each distance to a partition $M$.
At the end of reading of the whole data set we
have formed rings with respect to partitions, 
$\left\langle dN_p(M) \right\rangle_e$\footnote{The rings here are adjusted to include pairs with distances  
belonging to the
partition $M$ but not to the partitions $M+1$ or $M-1$.}. Then we
can find the total number of pairs corresponding to the
partition $M$ by adding the relevant rings as in
eq.~(\ref{eq:N(R)_e}):
\begin{equation}\label{eq:N(M)_e}
\left\langle N_p(M)\right\rangle_e = 
\sum_{l\geq M} \left\langle dN_p(l)\right\rangle_e  
=\left\langle N_p(M+1)\right\rangle_e + \left\langle dN_p(M)\right\rangle_e\;.
\end{equation}
Then, through eq.~(\ref{eq:Np_F2}), we have 
\begin{equation}\label{eq:F2_Np}
F_2(M)=\frac{M^d}{\left\langle N_{m} \right\rangle _e^2}
2 \left\langle N_p(M)\right\rangle_e\;.
\end{equation}
So, in the evaluation of $F_2$ with this technique,
called $F_{2,CI}$,
we utilize the distances $R$ between pairs, through
movable disks, as it is
done in $C$ and the relation $M(R)$ between partitions
and scales to form rings of partitions, $dN_p(M)$.
Thus, this procedure determines
 the central quantity which is the
pairs that correspond to a certain partition $N_p(M)$.
At conventional calculations of 2nd order factorial moments,
$F_{2,c}$, we estimate the number of pairs at a certain
partition directly by counting points
within fixed cells.
These calculation schemes of $F_2$ can be summarised as 
\begin{equation}\label{eq:F2CI-F2c}
R,\;M(R),\;
\left\langle dN_p(M)\right\rangle_e \stackrel{F_{2,CI}}{\longrightarrow}
\left\langle N_p(M)\right\rangle_e \stackrel{F_{2,c}}\longleftarrow
\sum\limits_c \left\langle {N(N - 1)} \right\rangle _e\;.
\end{equation}
The benefit we have by evaluating $F_2$ utilizing rings, is that we can achieve even more reduction in processing time compared to the conventional 
techniques, at least at the cases, as we shall see later on, where the conventional computing 
technique of $F_2$ is more time consuming than the relevant one of $C$.

As far as the error calculation is concerned, the total 
error of $F_2$, for the ring technique (using eq.~(\ref{eq:F2_Np})) or conventional manner (using eq.~(\ref{eq:Np_F2})), can be found to be
\begin{equation}\label{eq:deltaF_2}
\delta F_2(M)=F_2(M)
\sqrt{
\left[\frac{\delta \left\langle N_p(M)\right\rangle_e}
{\left\langle N_p(M)\right\rangle_e}\right]^2+
\left[2 \frac{\delta \left\langle N_m \right\rangle_e}
{\left\langle N_m \right\rangle_e}\right]^2
}\;,
\end{equation}
where the errors $\delta \left\langle N_p(M)\right\rangle_e$ 
are evaluated in a similar way to eq.~(\ref{eq:deltaN(R)}) and we have included the case of varying $N_m$ per event.
The only difference in the error calculation of $F_{2,CI}$
and $F_{2,c}$ comes from the different way the number
of pairs at a certain partition is evaluated according to 
scheme (\ref{eq:F2CI-F2c}).
We note that the errors (\ref{eq:deltaN(R)}) are the mean 
value errors and the errors (\ref{eq:deltaC}) and 
(\ref{eq:deltaF_2}) have been
calculated through error propagation method.
In error calculation throughout this paper we do
not use the bootstrap method. In the later, one forms
new data sets using the original set by reusing data points
repeatedly. For each new data set the standard error 
calculation is applied. The error calculation is carried
out as many times as the number of the formed data sets.
So, if there is benefit in time consumption in the
error evaluation for one data set with the ring
technique compared to conventional method, then this
benefit will be multiplied by a factor equal to the number of data sets
that will be processed in the bootstrap method.

Next, we come to the discussion on how we assign pairs to
scales or partitions. Firstly, in the $C$ calculation, to assign a pair to a ring of radius exactly $R$ (infinitely thin), we calculate the distance $S_D$ in the embedding space of 
dimension $d$, between the two points 1 and 2 of the pair,
using the $d$ coordinates $x$ of each point,
\begin{equation}\label{eq:R_F2}
R=S_D=\sqrt{\sum_{i=1}^d {\left(x_{1i}-x_{2i}\right)^2}}\;.
\end{equation}
For the $C$ calculation we divide the maximum available distance in the data $R_T$ in bins and we 
count them with the number $M_{CI}$, with
the zero bin corresponding to maximum distances and the maximum bin corresponding to minimum distances.
Since, we usually plot $C$ in logarithmic scales, we choose bins which appear equal in such scales. Therefore, we choose a relation 
between number of bin and scale like
\begin{equation}\label{eq:R_C}
R \cdot \tilde{A}_b^{M_{CI}}=R_T,
\end{equation}
where $\tilde{A}_b>1$, but close to unit. Increasing the number of bins will 
require to use values of $\tilde{A}_b$ closer to unit. 
In general, if we want to access results until a minimum scale $R_{\min}$
and divide the length between $R_{\min}$ and $R_T$ in $M_{CI,\max}$
bins, which will appear equal in a logarithmic axis, then we
have to choose $\tilde{A}_b$ as
\begin{equation}\label{eq:A_b}
\tilde{A}_b=\left(\frac{R_T}{R_{\min}}\right)^{1/M_{CI,\max}}\;.
\end{equation}
On the other hand, if we choose $\tilde{A}_b$ and $M_{CI,\max}$
independently, then the lower scale we will be able to access will be
\begin{equation}\label{eq:A_b}
R_{\min}=R_T \cdot \tilde{A}_b^{-M_{CI,\max}}\;.
\end{equation}
In eq.~(\ref{eq:R_C}) $M_{CI}$ is considered as a real number. To convert it to integer for bin purposes we can use the integer part, 
after solving for $M_{CI}$,
\begin{equation}\label{eq:M_CI-m_c}
M_{CI}=\left[\ln\left(\frac{R_T}{R}\right)\frac{1}{\ln(\tilde{A}_b)} +m_c \right],
\end{equation}
where the brackets indicate the integer part and $0 \le m_c < 1$, a parameter through which we can impose a small shift to the
scales that correspond to a specific bin.
In our calculations we shall take $m_c=0$.
The last equation assigns different scales within an interval $dR$ to a particular integer $M_{CI}$ and
so, our ring has now become finitely thin. Each pair with distance $s_D=R$ corresponds to a bin (ring) labelled by $M_{CI}$ according 
to eq.~(\ref{eq:M_CI-m_c}).
We, also, choose a maximum number of divisions 
$M_{CI,\max}$. 
Pair distances that lead to $M_{CI}>M_{CI,\max}$ are 
assigned to $M_{CI,\max}$.

Secondly, in the $F_2$ calculation, we can evaluate the scale $R=s_D$ of a pair and then we can find the
relevant $M$ according to eqs.~(\ref{eq:eq_spaces_compact}), which in that sense is a real number.
To convert it to integer we can use
\begin{equation}\label{eq:M-m_f}
M=\left[\frac{\beta_d \cdot R_w}{R} +m_f \right],
\end{equation}
where $0 \le m_f < 1$, a parameter though which we can impose a small shift to the scales that correspond to a specific bin. 
The value of this parameter has no effect for large values of $M$, however it does affect the low 
$M$ calculations. We shall choose it appropriately to match as much as possible the calculation 
of $F_2$ though the ring technique with the conventional estimation for the few estimation points
of low $M$.

In the following we probe the effectiveness of the new algorithm with respect to conventional techniques.
For this reason we perform calculations for the $F_2$ with the conventional method, which we shall
call $F_{2,c}$. In this algorithm we have improved the time consumption by placing the data points to
the specific grid cells they belong to for a particular partition $M$ and not the opposite 
(i.e.~investigate whether each cell contains any of the data points). 
Also, in this algorithm we do
not apply any grid averaging, so for every partition $M$ the calculation is carried out once. 
Had we done this averaging, the consumed time would be multiplied by the number of different 
grid locations we performed the calculation.
The calculations with our new algorithm for $F_2$ are
called $F_{2,CI}$, and in these cases we first perform correlation integral like calculations with the ring technique and then
map the results to $F_2$. We perform calculations for $C$ with the conventional method, which we 
call $C_{c}$, where for each scale $R$ we form disks of radius $R$ and search for the number of 
pairs that are enclosed within. The calculations with our new algorithm for $C$, where we first 
assign pairs to rings are called $C_{r}$. We measure the time consumption for the 
calculations with these techniques with respect to the number of calculation points, $N_D$ 
(total different partitions
$M$ in $F_2$, or total different scales $R$ in $C$), the number of events of the data set, $N_e$ and the number of
multiplicities $N_m$ per event (which corresponds to $\frac{1}{2} N_m(N_m-1)$ number of pairs per 
event). The $F_{2,CI}$ and $C_{r}$ calculations are carried out with estimation of errors and 
without errors. 
All measurements of time are performed in the same computing machine and for data sets we have used
uniform distributions in $d$-dimensional spaces. In every projection of the embedding space, the 
data points we have used are uniformly
distributed in the interval $[-5,5]$.  We, also, exclude from all measurements the
time needed for the output of the final results.

In Figs.~\ref{fig:MF-MC}-\ref{fig:TF-TC} we present time calculations for $F_2$ (in (a)) and $C$
(in (b)) as function of 
estimation
points $N_D$, number of events, $N_e$ and multiplicities per event, $N_m$, respectively,
while keeping the remaining parameters constant. 
The estimation points $N_D$ for $F_2$ are points that
correspond to all partitions up to a higher partition $M_{max}$, so $N_D=M_{max}$.
The retained
constant parameters for every case are so chosen in order to have enough measurements for all techniques
with times $<\sim 300$ sec.
In each graph we also depict curves appearing as straight lines in the logarithmic plots, which approximate the recorded data for large values of the parameter
under investigation. Our aim is to reveal the exponent of the parameter which influences the
calculation time. The relation that connects time with the parameter and the
accompanying exponent are depicted on the graphs.

Fig.~\ref{fig:MF-MC}(a) 
shows that for high values of $N_D$, time for $F_{2,c}$, for constant $N_e$ and $N_m$, grows
proportionally to $N_D^{d+1}$. On the contrary, $F_{2,CI}$, for constant $N_e$ and $N_m$, grows
proportionally to $N_D$ with error estimation and is almost independent of $N_D$ without error 
estimation. 
This reveals the significant advantage which our new algorithm offers when we want to
carry out calculations
for high number of estimation points, that is when we want to reach low scales
$R$, 
or high partitions $M$.
The advantage becomes more significant as the dimension of the 
embedding space, $d$, increases.
From Fig.~\ref{fig:MF-MC}(b) 
it is evident that for large values of $N_D$, time for both $C_{c}$ and $C_{r}$, for constant $N_e$ and 
$N_m$, grows proportionally to $N_D$. However, time values are lower in the case of the ring technique with the
effect being more prominent as the dimension $d$ increases. Also, comparing the conventional
techniques for $F_2$ and $C$ (Figs.~\ref{fig:MF-MC}(a) and (b) respectively) 
we find that $C$ is more advantageous, as far as the estimation points are concerned,
since it grows proportionally  \linebreak

\vspace{0.cm}
\begin{figure}[H]
\begin{minipage}[]{1.\textwidth}
\centering
\hspace{-0.2cm}\includegraphics[scale=0.43,trim=0.cm 3.2cm 0.cm 1.3cm,angle=0]{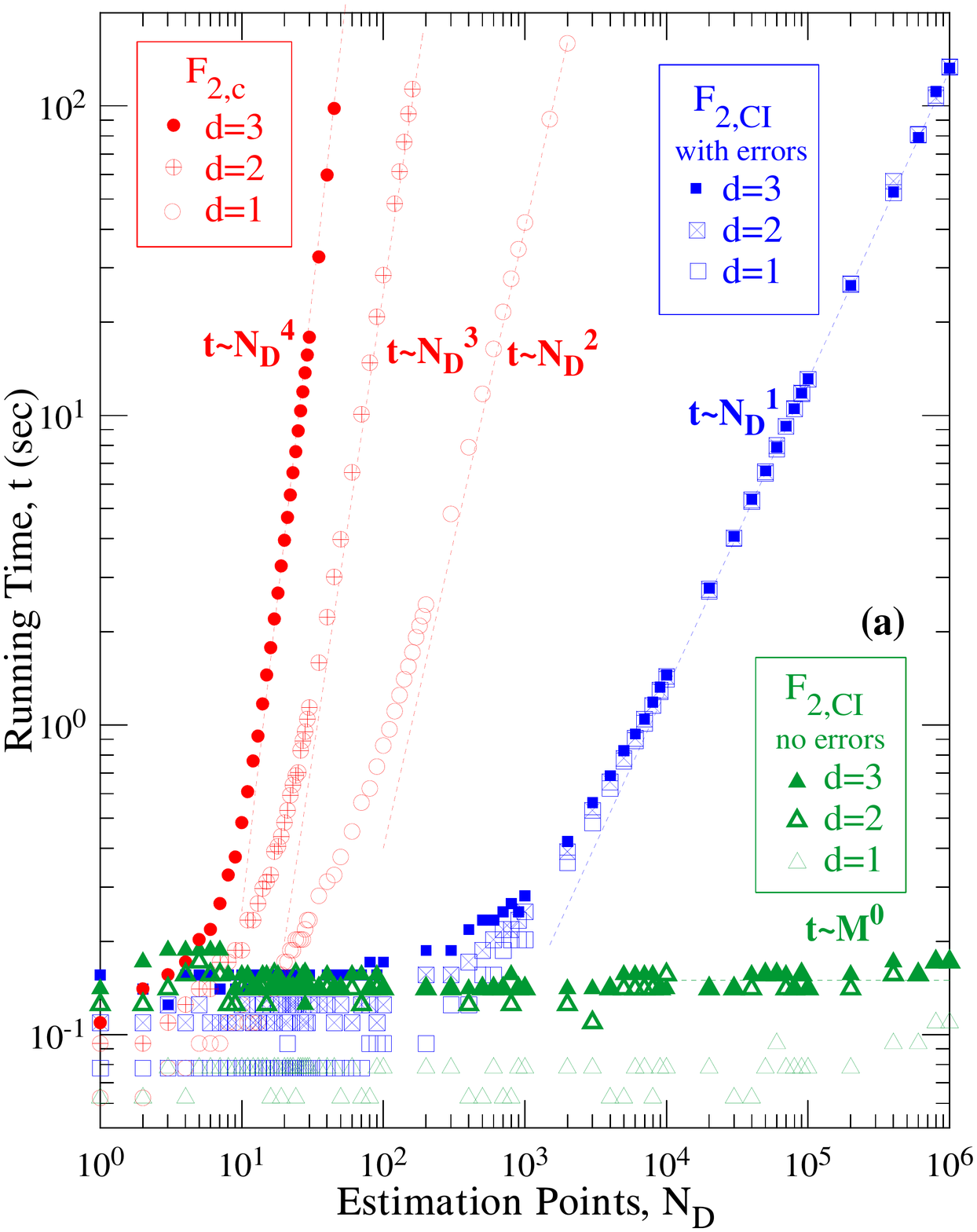}
\hspace{-0.3cm}\includegraphics[scale=0.43,trim=0.cm 2.9cm 0.cm 1.3cm,angle=0]{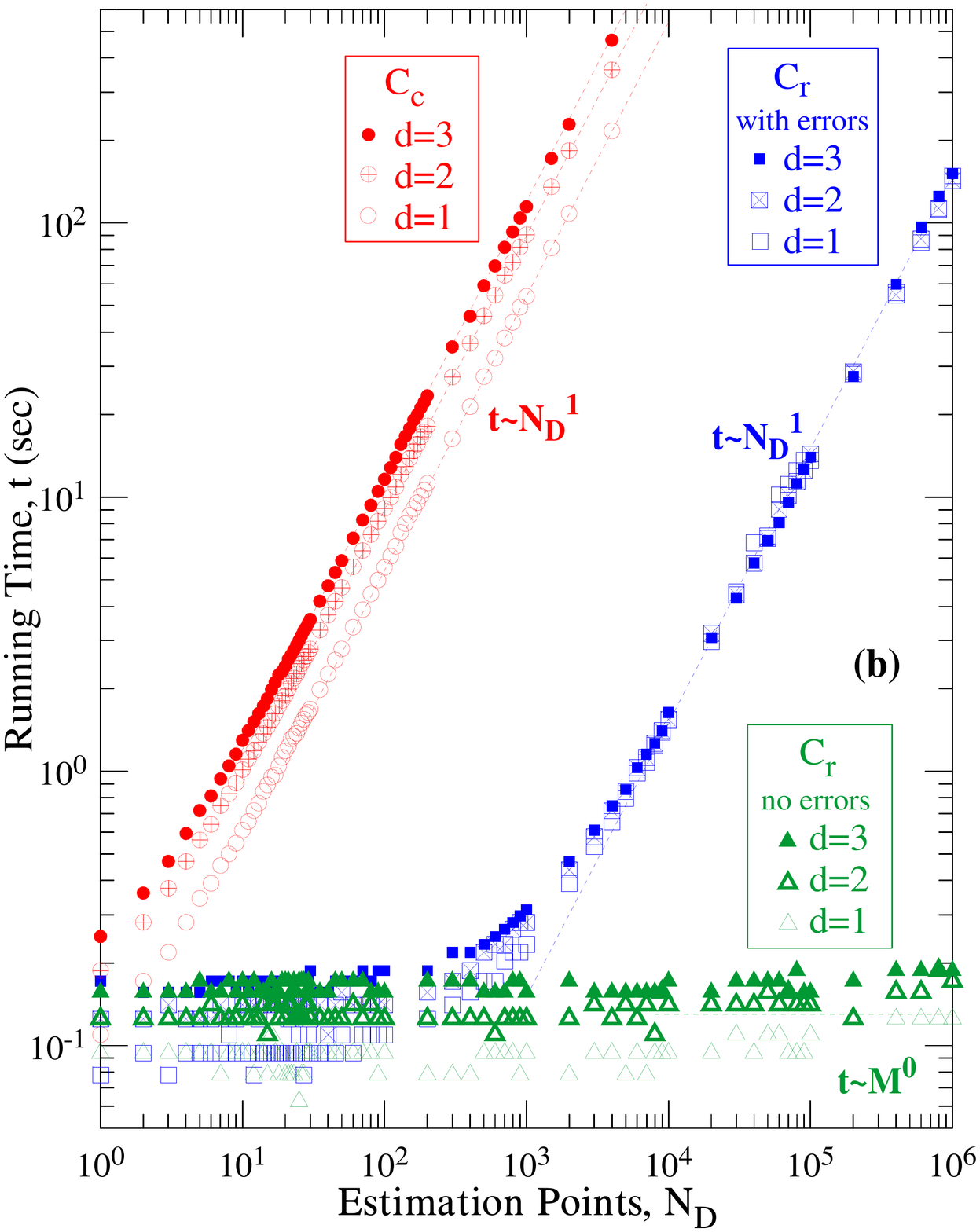}\\
\vspace{-0.2cm}
\caption{\label{fig:MF-MC} {\small The running time for $F_2$ in (a) and $C$ in (b) and for different 
techniques as function of the estimation points $N_D$, for fixed $N_m=$10 and $N_e=$10000.}}
\end{minipage}
\vspace{-0.0cm}
\end{figure}

\vspace{-0.cm}
\begin{figure}[H]
\begin{minipage}[]{1.\textwidth}
\centering
\hspace{-0.2cm}\includegraphics[scale=0.43,trim=0.cm 2.2cm 0.cm 2.3cm,angle=0]{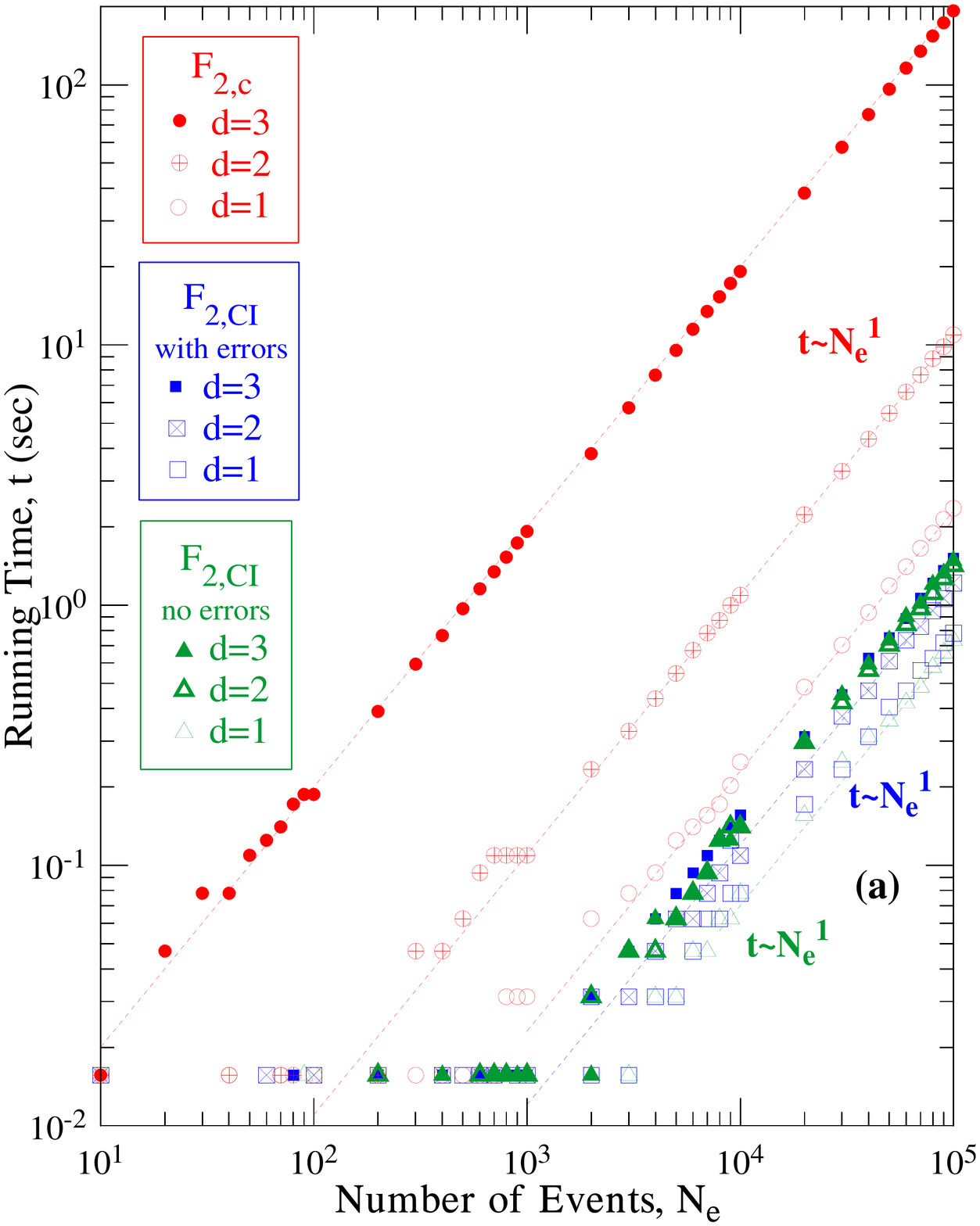}
\hspace{-0.3cm}\includegraphics[scale=0.44,trim=0.cm 2.5cm 1.cm 2.3cm,angle=0]{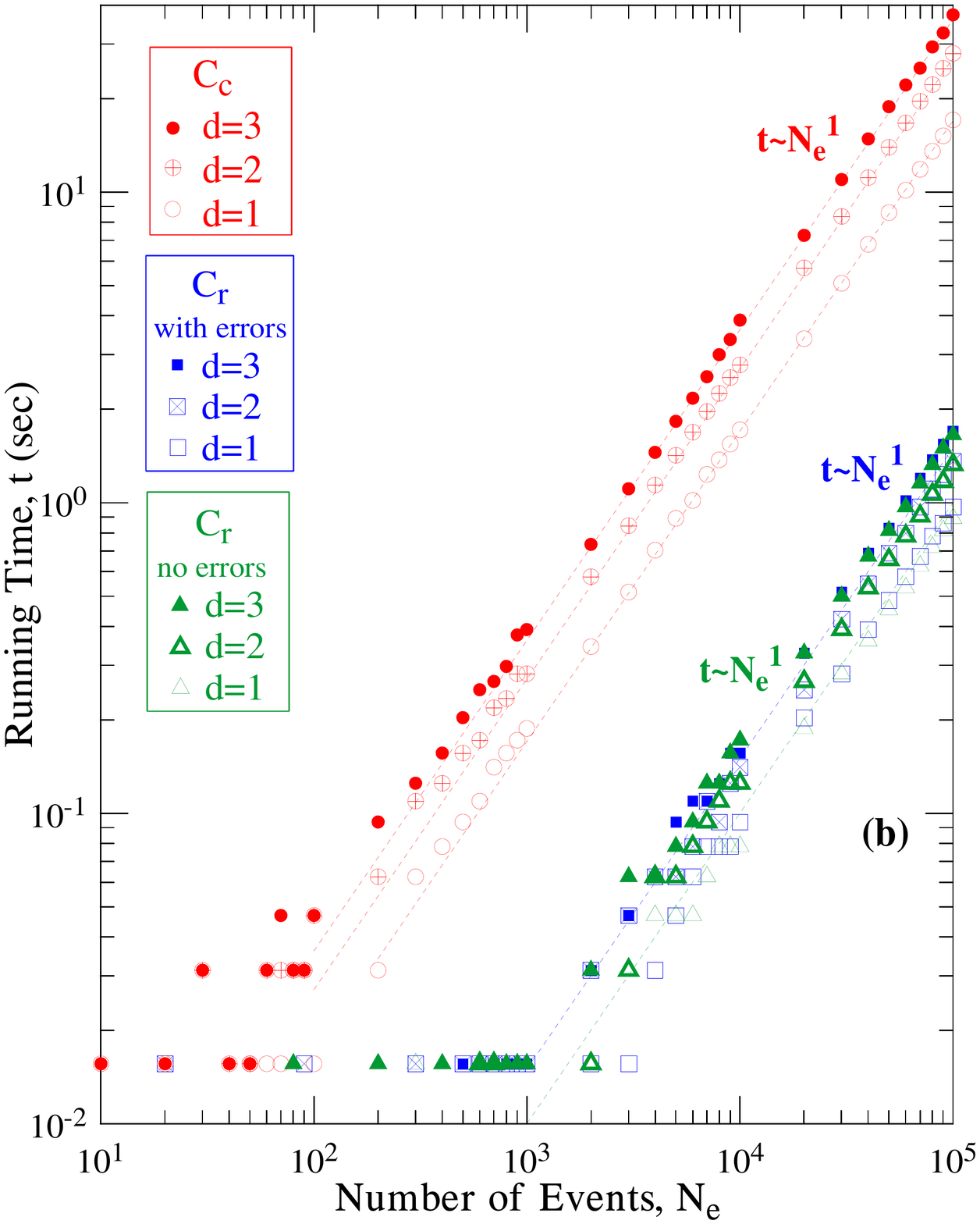}\\
\vspace{-0.2cm}
\caption{\label{fig:EF-EC} {\small The running time for $F_2$ in (a) and $C$ in (b) and for different 
techniques as function of the number of events $N_e$, for fixed estimation points $N_D=$30 and
$N_m=$10.}}
\end{minipage}
\vspace{0.1cm}
\end{figure}


\vspace{-0.3cm}
\begin{figure}[H]
\begin{minipage}[]{1.\textwidth}
\vspace{0.0cm}
\centering
\hspace{-0.2cm}\includegraphics[scale=0.43,trim=0.cm 2.5cm 0.cm 1.9cm,angle=0]{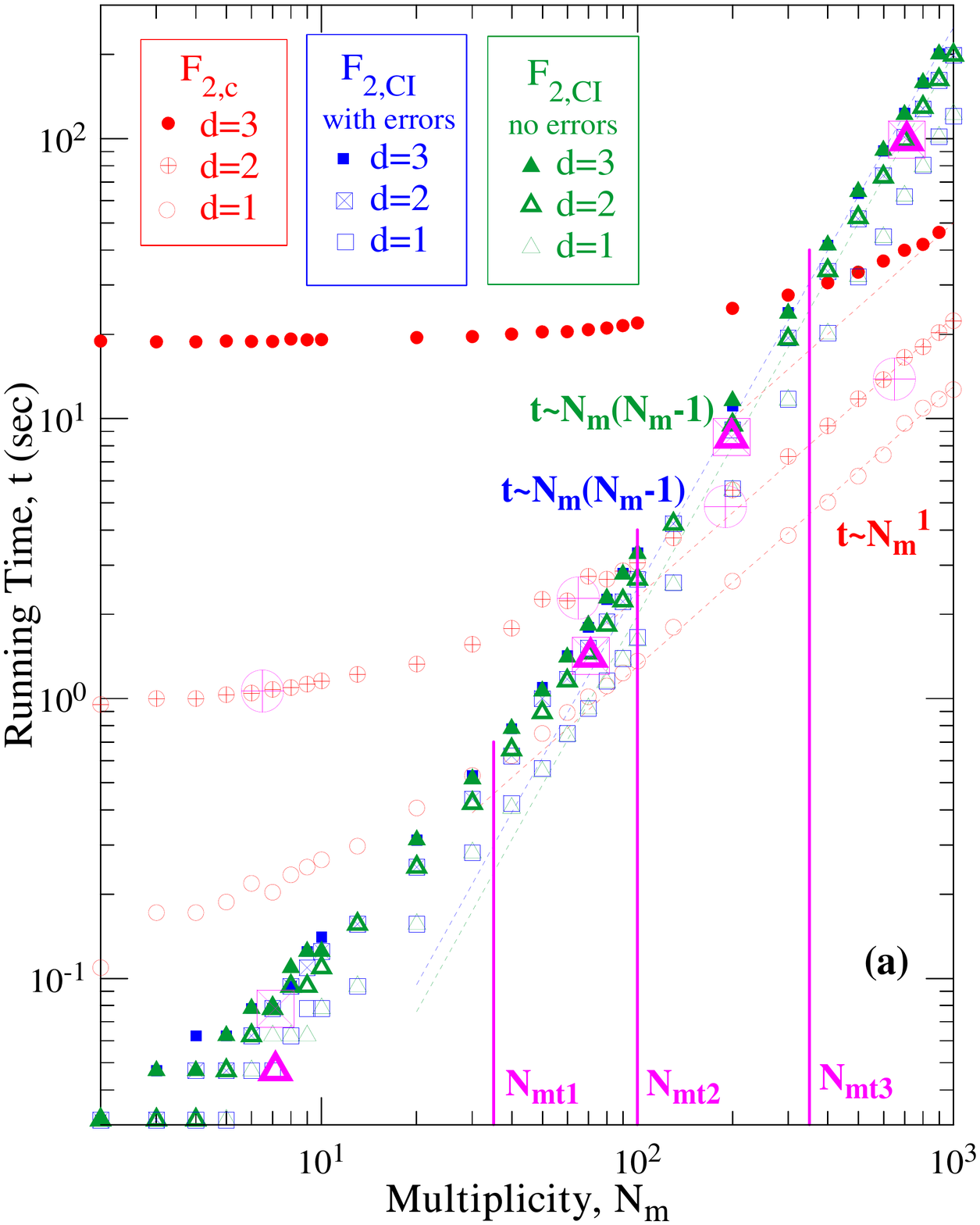}
\hspace{-0.3cm}\includegraphics[scale=0.43,trim=0.cm 2.2cm 0.cm 1.9cm,angle=0]{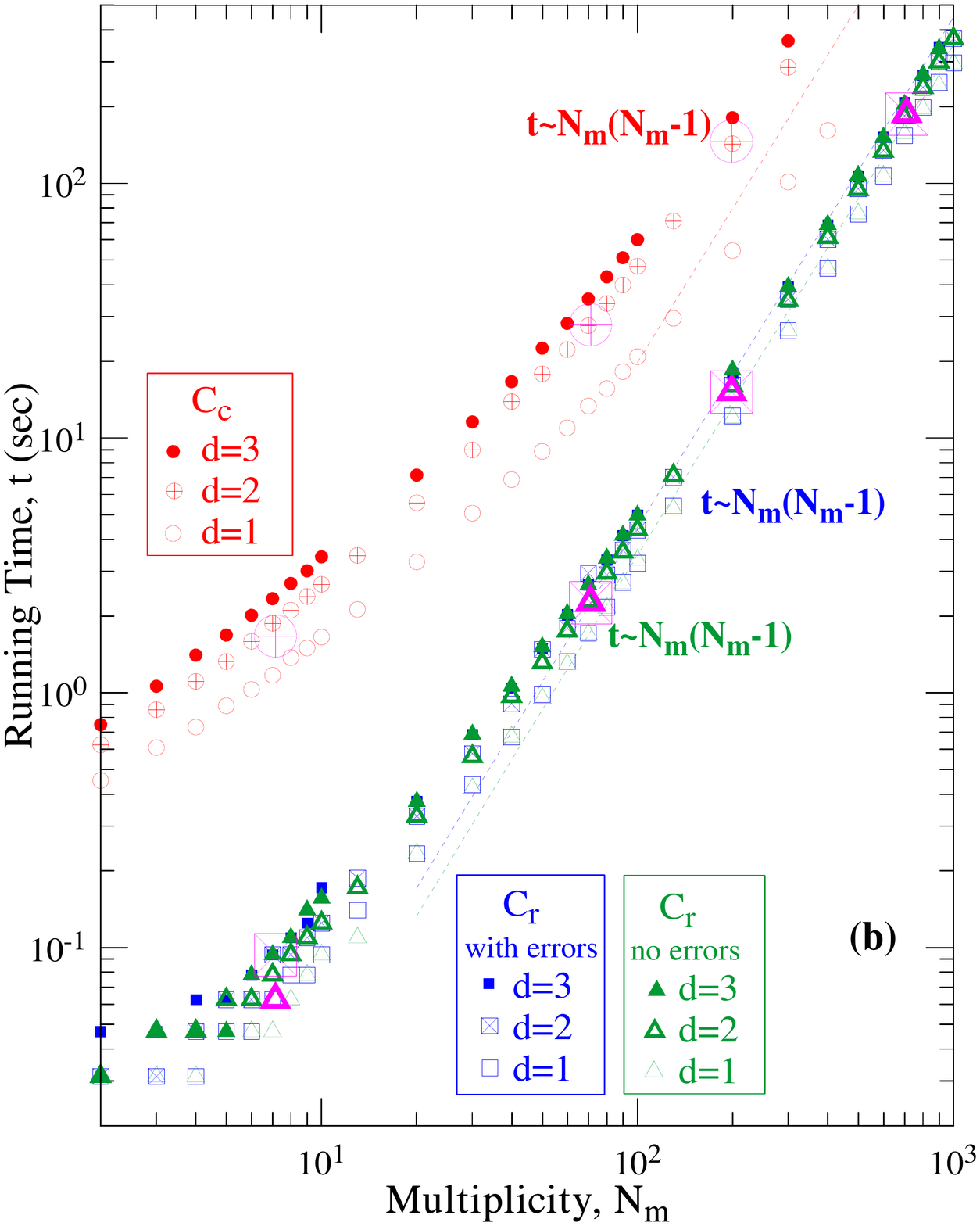}\\
\vspace{-0.2cm}
\caption{\label{fig:TF-TC} {\small The running time for $F_2$ in (a) and $C$ in (b) and for different 
techniques as function of the multiplicity $N_m$, for fixed estimation points $N_D=$30 and
$N_e=$10000. Points with greater size symbols
correspond to varying $N_m$ cases.}}
\end{minipage}
\vspace{-3cm}
\end{figure}

\clearpage

\noindent
to $N_D$ raised to a lower exponent compared to $F_2$.
The $F_2-C$ mapping offers the opportunity to use
$C$-based calculations to evaluate $F_2$
with the accompanying reduction in time consumption.

In Fig.~\ref{fig:EF-EC}(a)-(b) 
we see that for high values of number of events $N_e$, at constant estimation points $N_D$ and multiplicities $N_m$, time for all 
techniques grows proportionally to $N_e$.

In Fig.~\ref{fig:TF-TC}(a)-(b) we see how calculation time varies with respect of the multiplicities
per event, $N_m$, for constant estimation points $N_D$ and events $N_e$.
For high values of $N_m$ we expect that time for $F_{2,c}$ grows proportionally to $N_m$.
However, the conventional method of calculation of the correlation integral, $C_c$, as well as,
all methods of calculation with 
the ring technique, $F_{2,CI}$ and
$C_r$ grow proportionally to 
the number of pairs $\frac{1}{2}N_m(N_m-1)$.
Thus, it is inevitable that at some
 value of $N_m$ the conventional $F_{2,c}$ will become more 
effective than the ring technique. The values of $N_m$ 
that this occurs are represented in the
graphs by values $N_{mtd}$, where $d$ is the space dimension. $N_{mtd}$ increases with $d$. 
For the 
parameters depicted in the graph, $N_{mt1}$ (for $d=1$) is of the order of magnitude of the estimation 
points, $N_{mt1} \simeq N_D$, while as $d$ increases $N_{mtd}$ is pushed at even higher values than the 
estimation points.
 
In our calculations we assumed that $N_m$ is fixed for all events.
Since time consumption is, in all cases, proportional to the number
of events, we prove in the Appendix that, when $N_m$ varies, the calculation time
changes by replacing any function of $N_m$ by its average value with respect to
the events. Additionally, to test the validity of this
finding, we measure the time for processing datasets of events with
varying $N_m$. Specifically, we prepare files with total
$N_e=$10000, subdivided to 10 clusters of 1000 events, with each
cluster containing events with fixed multiplicity 
$N_{m,j}=N_{m,i}+jN_s$, $j=0,\ldots,9$. We choose appropriately the initial 
multiplicity $N_{m,i}$ and the step $N_s$ in order to have total computational 
times within the range of Fig.~\ref{fig:TF-TC}. We use again
uniform distribution in 2-dimensional embedding space in a
rectangle $[-5,5]\times[-5,5]$. We only perform calculations for
$d=2$ and for $N_D=$ 30 and the resulting points are placed on Fig.~\ref{fig:TF-TC},
using the same symbol with the corresponding fixed $N_m$ results,
but of greater size. To place a point on the graph, which has
as horizontal axis the multiplicity $N_m$, we have to evaluate
which is the appropriate single value $N_m$ in the case of
varying $N_m$. 
In the case of $F_{2,c}$ 
which for constant multiplicity is proportional to $N_m$, this is just
$\left\langle N_m \right\rangle_e $.
For the rest of cases we 
calculate the average number of pairs
with respect to the number of events,
$\left\langle \frac{1}{2} N_m(N_m -1) \right\rangle_e$ $ \equiv w$.
Then we solve the equation $\frac{1}{2} x(x -1) =w$ and the value
$x$ is the one that corresponds to $N_m$ which appears to the
axes of Fig.~\ref{fig:TF-TC}.
We observe that the resulting points fall on the same curve 
which is formed by the fixed multiplicity cases.

In conclusion, the calculation time for $F_2$ and $C$ with the various techniques for large values of
the depending parameters grows like the forms listed in Table \ref{tab:t}.


\begin{table}[H]
\centering
\begin{tabular}{|c|c|c|} \hline
Time & Grows like ($\propto$) & Grows like ($\propto$)\\ 
     & (fixed $N_m$) & (varying $N_m$) \\ \hline
$t(F_{2,c})$ & $ N_D^{d+1}  N_e  N_m$ & 
$ N_D^{d+1}  N_e  \left\langle N_m \right\rangle_e $ \\
$t(F_{2,CI}){\rm \;(with\;errors)} $ & $ N_D  N_e  \frac{1}{2} N_m(N_m -1)$ & 
$ N_D  N_e  \left\langle \frac{1}{2} N_m(N_m -1) \right\rangle_e $\\
$t(F_{2,CI}){\rm \;(no\;errors)}$ & $ N_e  \frac{1}{2} N_m(N_m -1) $ 
& $ N_e  \left\langle \frac{1}{2} N_m(N_m -1) \right\rangle_e $ \\
$t(C_c)$ & $ N_D  N_e \frac{1}{2} N_m(N_m -1) $ & 
$ N_D  N_e \left\langle \frac{1}{2} N_m(N_m -1) \right\rangle_e $ \\
$t(C_r){\rm \;(with\;errors)}$ & $ N_D  N_e \frac{1}{2} N_m(N_m -1) $ 
& $ N_D  N_e \left\langle \frac{1}{2} N_m(N_m -1) \right\rangle_e $ \\
$t(C_r){\rm \;(no\;errors)}$ & $ N_e \frac{1}{2} N_m(N_m -1) $ &
 $ N_e \left\langle \frac{1}{2} N_m(N_m -1) \right\rangle_e $ \\ \hline  
\end{tabular}
\vspace{-0.0cm}
\caption{\label{tab:t} {\small The dependence on the involved parameters, $N_D$, $N_e$ and $N_m$ (for large values),
of the computation times.}}
\end{table}

From Table \ref{tab:t} we see that, if we want to decide whether 
 the conventional or the
ring method offers faster calculation of $F_2$, we have to compare $N_D^d$ with $\frac{1}{2} (N_m -1)$ in the presence of 
error calculation, or $N_D^{d+1}$ with $\frac{1}{2} (N_m -1)$ in the absence of 
error calculation. The exact result, of course, will
depend of the accompanying factor of these equations.
In the correlation integral case we always have advantage using the ring technique compared to the
conventional one. This advantage becomes more significant as the space dimension $d$ increases
and is more profound in the absence of error calculation. However, the advantage in time consumption
is expected to smear as the multiplicity per event increases.
If we want to investigate what goes on at low scales $R$, or high partitions $M$, we must
have high number of estimation points. So, it becomes apparent
that the maximum number of multiplicities up to which the ring technique retains its effectiveness 
is pushed
to high values, thus, making this technique an indispensable tool in such cases.

The source code of our program in FORTRAN 90 is provided as supplementary material to this paper, containing within
explanatory comments for setting the necessary parameters for a specific result.

\section{Applications}\label{sec:app}

\subsection{Data Analysis}\label{subsection:data}

We shall apply the correspondence between $F_2$ and $C$ in various data sets. 
Neglecting boundary phenomena for high scales $R$, $C$ can usually be approximated by the form 
\begin{equation}\label{eq:C-A}
C(R) \simeq A R^{d_F}\;,
\end{equation}
in accordance with eq.~(\ref{eq:CI_propto}).
Then, using eq.~(\ref{eq:F2-C}), we can approximate $F_2$ as
\begin{equation}\label{eq:F2-A}
F_2(M^d) \simeq
a_m \left(\beta_d R_w\right)^{d_F} A \left( M^d \right)^{1-\frac{d_F}{d}}\;,
\end{equation}
which is in agreement with eq.~(\ref{eq:F2_propto}).
As it is seen from the last two equations, the exponent of the
correlation integral reveals directly the dimension of
the data set under investigation, $d_F$, whilst $F_2$
shows indirectly this dimension through an exponent of the form
$1-\frac{d_F}{d}$. Both exponents of $C$ and $F_2$ can be
identified as the slopes in the corresponding logarithmic
plots.
As an example, in \cite{Crit-Opal} it is shown that the critical opalescence in QCD
matter is revealed in transverse momentum space as a
power-law behaviour in $q$-order factorial moments 
$F_q \sim (M^2)^{s_q}$, with 
$s_q=(q-1)\left(1-\frac{\tilde{d}_F}{2}\right)$ and $\tilde{d}_F=\frac{1}{3}$.
Thus, while the phenomenon of critical opalescence is expected to appear with a
slope in $F_2$ equal to $s_2=\frac{5}{6}$, the slope of $C$ will be directly
equal to the isotherm critical exponent of QCD, $\tilde{d}_F=\frac{1}{3}$.

We proceed by dealing with sets with different structures. 
We shall analyse sets of points which follow the uniform 
distribution in a region of volume (measure) $V_d$ of a space with dimension $d$. 
These points have constant probability
density everywhere in this region, which is
\begin{equation}{\label{eq:uniform}}
p(x) = \left( V_d \right)^{-1}\;.
\end{equation}
Here we shall present results for embedding spaces of one and three dimensions.
We shall, also, analyse a L\'{e}vy set \cite{Levy} in one dimension. 
This is produced with steps which follow the probability distribution
\begin{equation}{\label{eq:Levy}}
p(x) = \nu b^\nu \left[ 1 - {\left( \frac{b}{x_c} \right)}^v \right]^{-1}
x^{-1-\nu}\;,
\end{equation}
with $\nu=1/3$ and $x$ taking values between $b=10^{-2}$ and $x_c=10^5$. 
The $n-$th point in the set of $N_m$ multiplicities is the $n$-th successive step, with the first step
starting from the origin 0. Before taking each step, apart from the size of the step determined by 
eq.~(\ref{eq:Levy}), it is decided whether to move forward or backward using a uniform probability.
The increased number of multiplicities, $N_m=250$, is needed in order
to produce an almost linear part of the $C$-curve in the logarithmic plot.
In embedding space of two dimensions, we shall analyse sets of points taken from the Henon \cite{Hen}
and the Ikeda attractor \cite{Ike}.
In embedding space of three dimensions, we shall analyse a Lorentz set which is produced 
from points 
taken from the Lorentz attractor \cite{Lor1,Lor2} with parameters
$\rho=28$, $\sigma=10$ and $\beta=8/3$.

In graphs \ref{fig:1_uniform}-\ref{fig:3_lorentz} we present calculations for the factorial moments
$F_2$ in (a) and for the correlation integral $C$ in (b). We present examples for all three dimensions 
$d$
of the embedding space. The calculations for $F_2$ are carried out
using the correspondence to the correlation integral, $F_{2,CI}$ (open circles). The
calculations for $C$, $C_r$, as well as those for $F_2$ are carried out using the ring 
technique
(open circles). 
For comparison, we, also, show results for $F_2$ through the 
conventional grid technique, $F_{2,c}$, in (a) and their correspondence to $C$, $C_{F_2}$, in (b)
(open rectangles). These are limited to fewer estimation points due to the higher need in computation
time. 
In order to match $F_{2,CI}$ with $F_{2,c}$ for the few low values of $M$ we set in
eq.~(\ref{eq:M-m_f}) $m_f=0.27$ in the uniform cases for $d=1$ and $d=3$. In all other cases $m_f=0$.
In graphs (b) we present a curve, shown as straight line in the logarithmic plots,
which has the form of 
eq.~(\ref{eq:C-A}) and which approximates $C(R)$ away from the higher scales which are available in 
the data set. In all cases, but the uniform sets, this 
curve is produced by a fit in the part of the $C_r$-curve which is enclosed within the two slashed
lines.
In the two uniform cases (Figs.~\ref{fig:1_uniform}(b) and \ref{fig:3_uniform}(b)) the curve is 
produced using the exact theoretical value of the set. The curve 
shown as straight line in the logarithmic plots (a) is 
eq.~(\ref{eq:F2-A}) for the value $A$ given in graphs (b).
The exact form of this equation is
depicted on the graphs for each case. From the fit performed on the $C_r$-curve
(Figs.~(b)) we extract the fractal 
dimension\footnote{To have a common description for all data sets we use
throughout this paper the
term ``fractal'' dimension for all cases. This includes the uniform distribution, where the dimension is
identical to the embedding space dimension.
This, however, does not imply that the uniform sets exhibit a non-trivial
behaviour, as scales change.} $d_{F,CI}$ with the use of 
$C$. We, also, perform a fit 
to the part of $F_{2,CI}$-curve between the two slashed lines in graphs (a), to extract again the fractal dimension, $d_{F,F_2}$
from the $F_2$ curve. The slashed lines between graphs
(a) and (b) are connected trough eqs.~(\ref{eq:eq_spaces_compact}).
Our results are listed in Table \ref{tab:d_F} for all cases, along with theoretical values,
$d_{F,th}$, for the fractal dimension.

Also, it is interesting to observe Fig.~\ref{fig:1_uniform}, where $F_2$ is limited within a more constrained
interval of values (since it is expected to have zero slop) and the magnitude of statistical fluctuations
is shown more clearly. We see that the calculations of 
$F_2$ through
the correlation integral experience lower statistical fluctuations compared to the grid technique.

Another interesting observation can be drawn from Fig.
\ref{fig:1_Levy}, where we
see that the $F_2$ conventional calculations are divided in two separate
sets which correspond to odd and even partitions (i.e. $M$ is an odd or
even integer, respectively). The L\'{e}vy distribution is produced, in this
case, with the first step in every event starting from the same point,
the origin of the axis. As a result, the one-particle distribution, i.e.
the probability of having a particle at a certain interval, exhibits an
extremely sharp peek at the origin.
The working window is placed so that the origin
is at the center of this window. The odd partitions systematically contain a cell
enclosing the center of the peek which counts together points 
existing at both sides of this peek. But the even partitions systematically split the top of
the peek to different cells and so they count fewer points. As $M$ 
progresses to higher partitions, or lower scales, the effect is smoothed
out, as the size of the cells becomes smaller that the average width of 
the distribution and the difference in counting between odd and even
partitions diminishes.  This 
situation, of course, can be remedied by applying grid averaging, at the
cost of considerable increase of time processing of data, which
we do not apply here. 
On the contrary, the $F_2$ calculated using correlation integral like calculations is not 
affected by the location of the cells at each partition. Since it is 
using moving disks, it is counting all the pairs 
corresponding at a 
certain scale, which is the radius of the disk.
In this L\'{e}vy case we had 
full control
on where to \linebreak

\vspace{-1cm}
\begin{figure}[H]
\begin{minipage}[]{1.\textwidth}
\centering
\includegraphics[scale=1.18,trim=7.8cm 11.6cm 7.8cm 0.cm,angle=0]{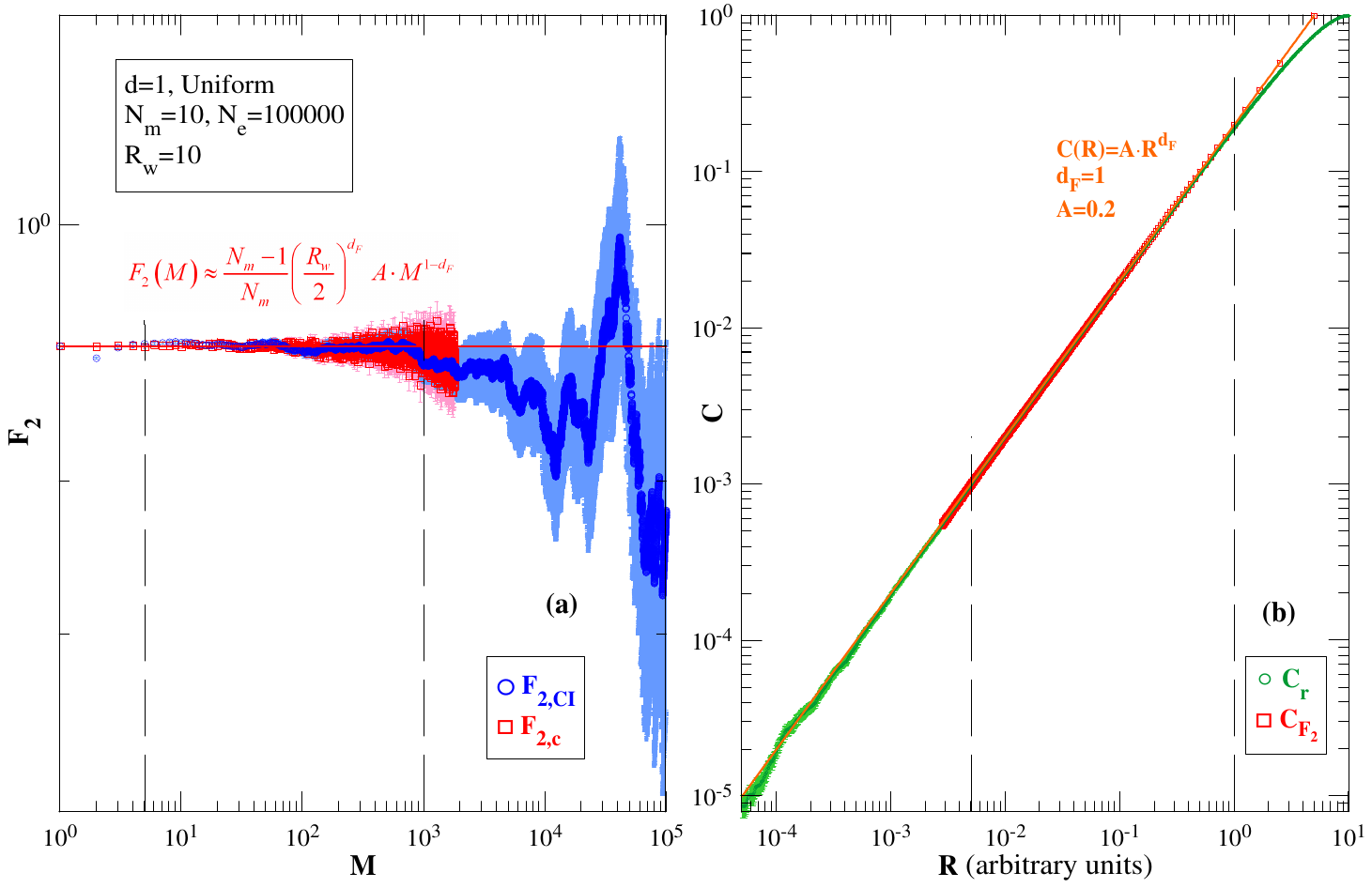}\\
\caption{\label{fig:1_uniform} {\small Analysis of a Uniform data set of $N_e=100000$ events and
multiplicity $N_m=10$ in an embedding space of $d=1$ dimension.}}
\end{minipage}
\vspace{-0.5cm}
\end{figure}

\begin{figure}[H]
\vspace{-0.cm}
\begin{minipage}[]{1.\textwidth}
\centering
\hspace{-0.2cm}\includegraphics[scale=0.7,trim=5.2cm 5.2cm 5.2cm 0.cm,angle=0]{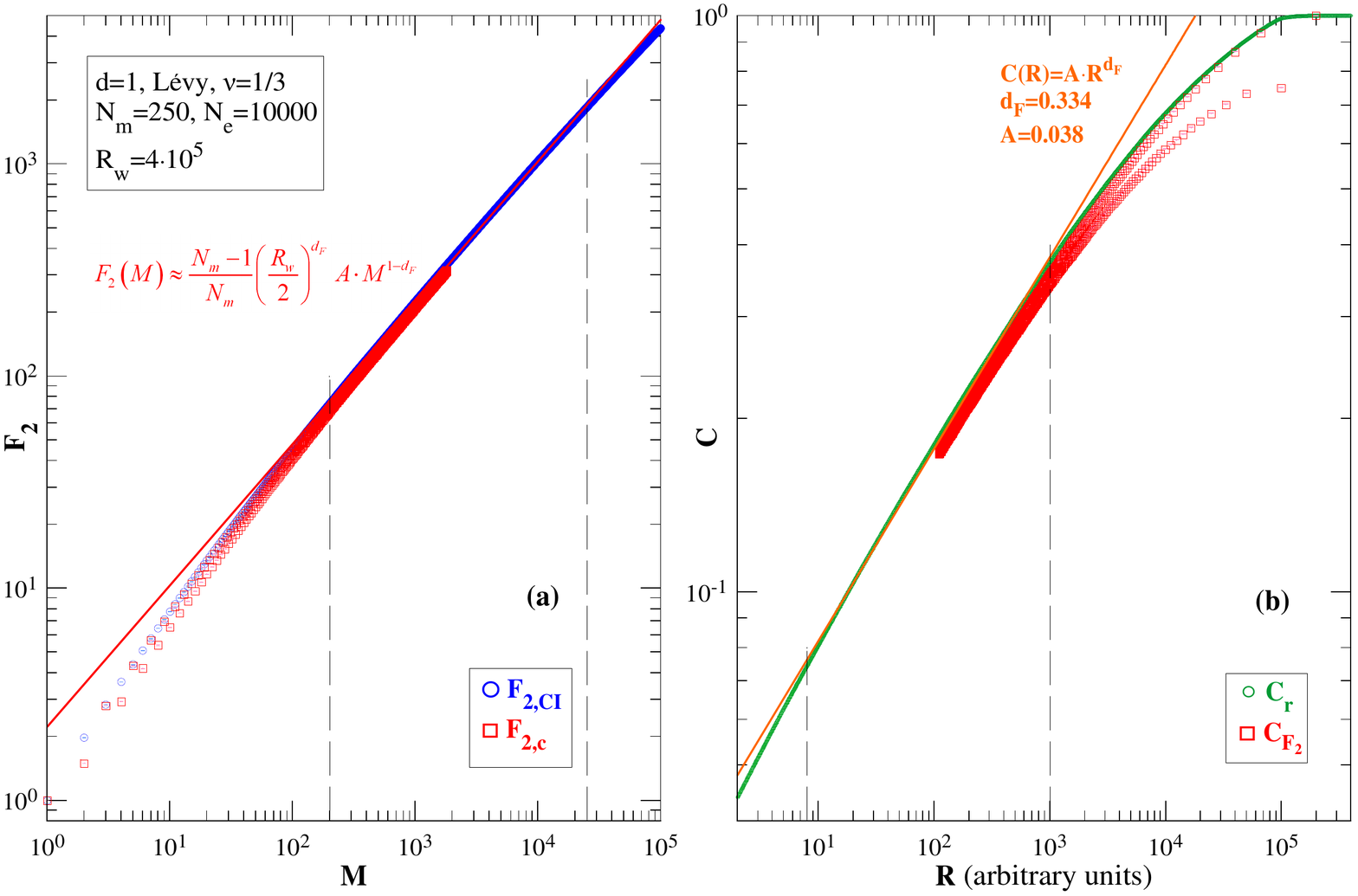}\\
\caption{\label{fig:1_Levy} {\small Analysis of a L\'{e}vy data set of $N_e=10000$ events and
multiplicity $N_m=250$ in an embedding space of $d=1$ dimension.}}
\end{minipage}
\vspace{-4.cm}
\end{figure}

\clearpage

\vspace{-1cm}
\begin{figure}[H]
\begin{minipage}[]{1.\textwidth}
\centering
\includegraphics[scale=0.59,trim=2.cm 2.2cm 2.cm 0.cm,angle=0]{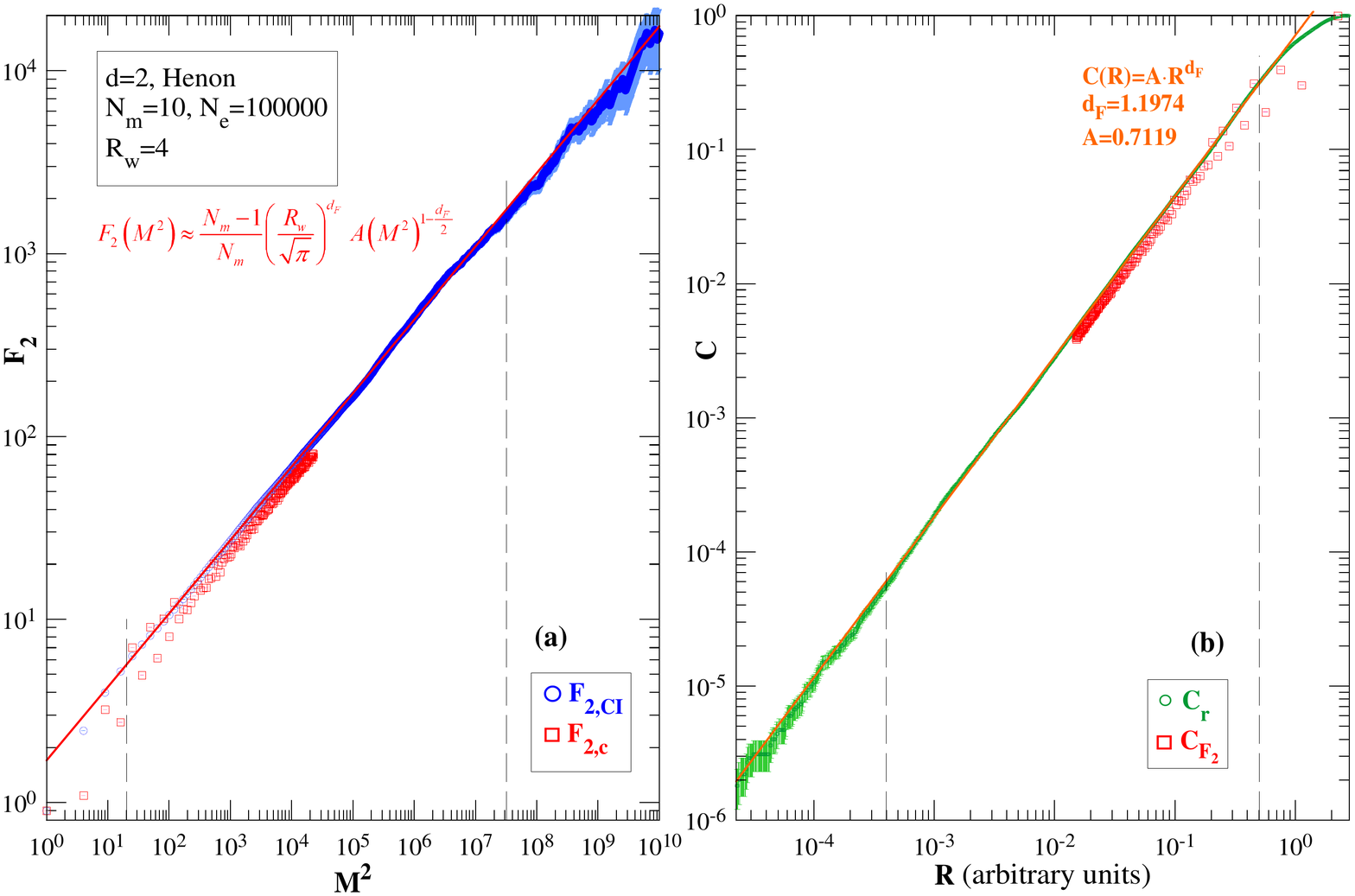}\\
\caption{\label{fig:2_henon} {\small Analysis of a Henon data set of $N_e=100000$ events and
multiplicity $N_m=10$ in an embedding space of $d=2$ dimensions.}}
\end{minipage}
\vspace{-0.5cm}
\end{figure}

\begin{figure}[H]
\begin{minipage}[]{1.\textwidth}
\centering
\includegraphics[scale=0.71,trim=4.3cm 5.5cm 4.3cm 0.cm,angle=0]{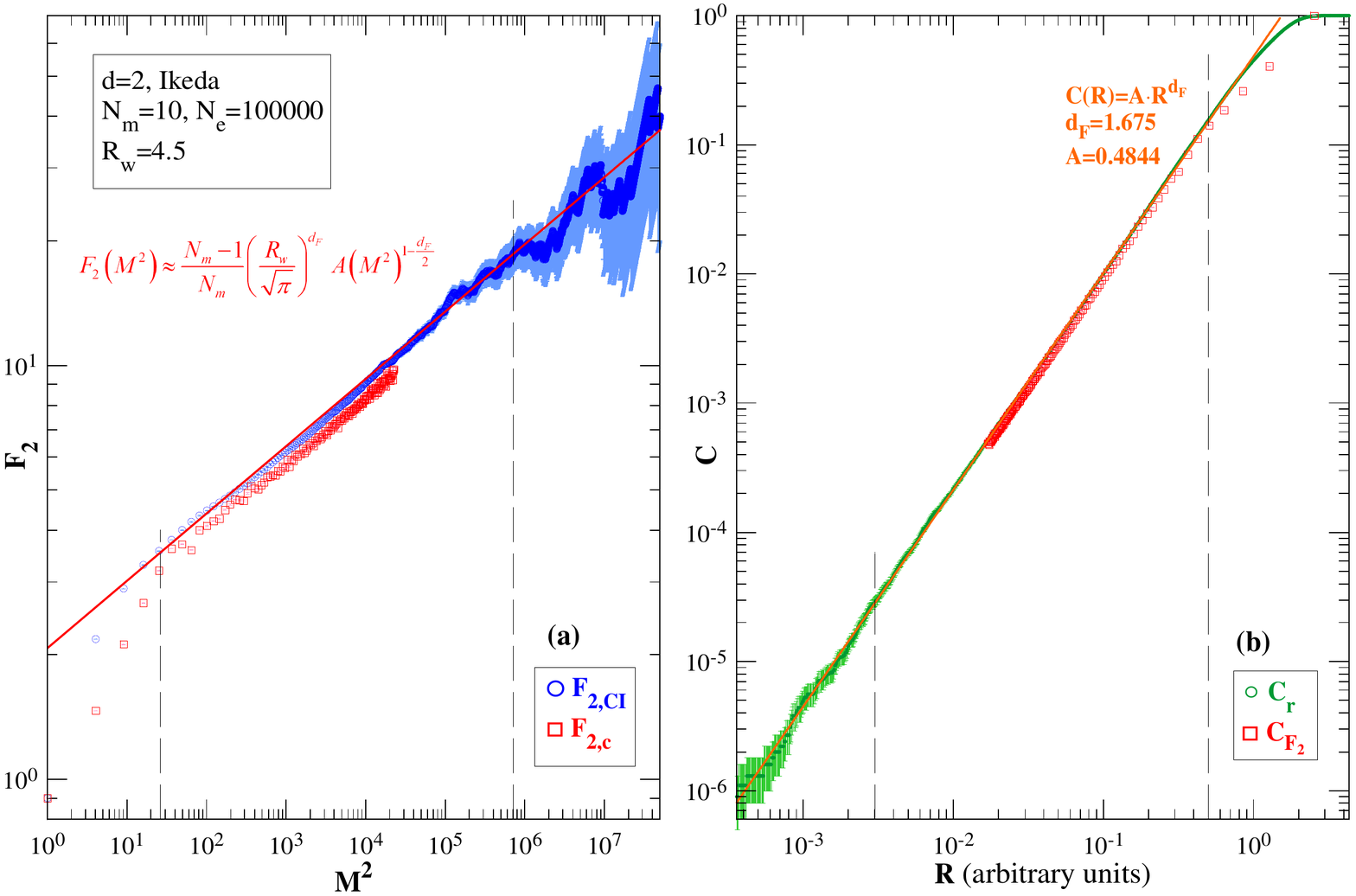}\\
\caption{\label{fig:2_ikeda} {\small Analysis of an Ikeda data set of $N_e=100000$ events and
multiplicity $N_m=10$ in an embedding space of $d=2$ dimensions.}}
\end{minipage}
\vspace{-0.cm}
\end{figure}

\clearpage

\vspace{-1cm}
\begin{figure}[]
\begin{minipage}[]{1.\textwidth}
\centering
%
\includegraphics[scale=0.82,trim=5.7cm 7.5cm 5.7cm 0.cm,angle=0]{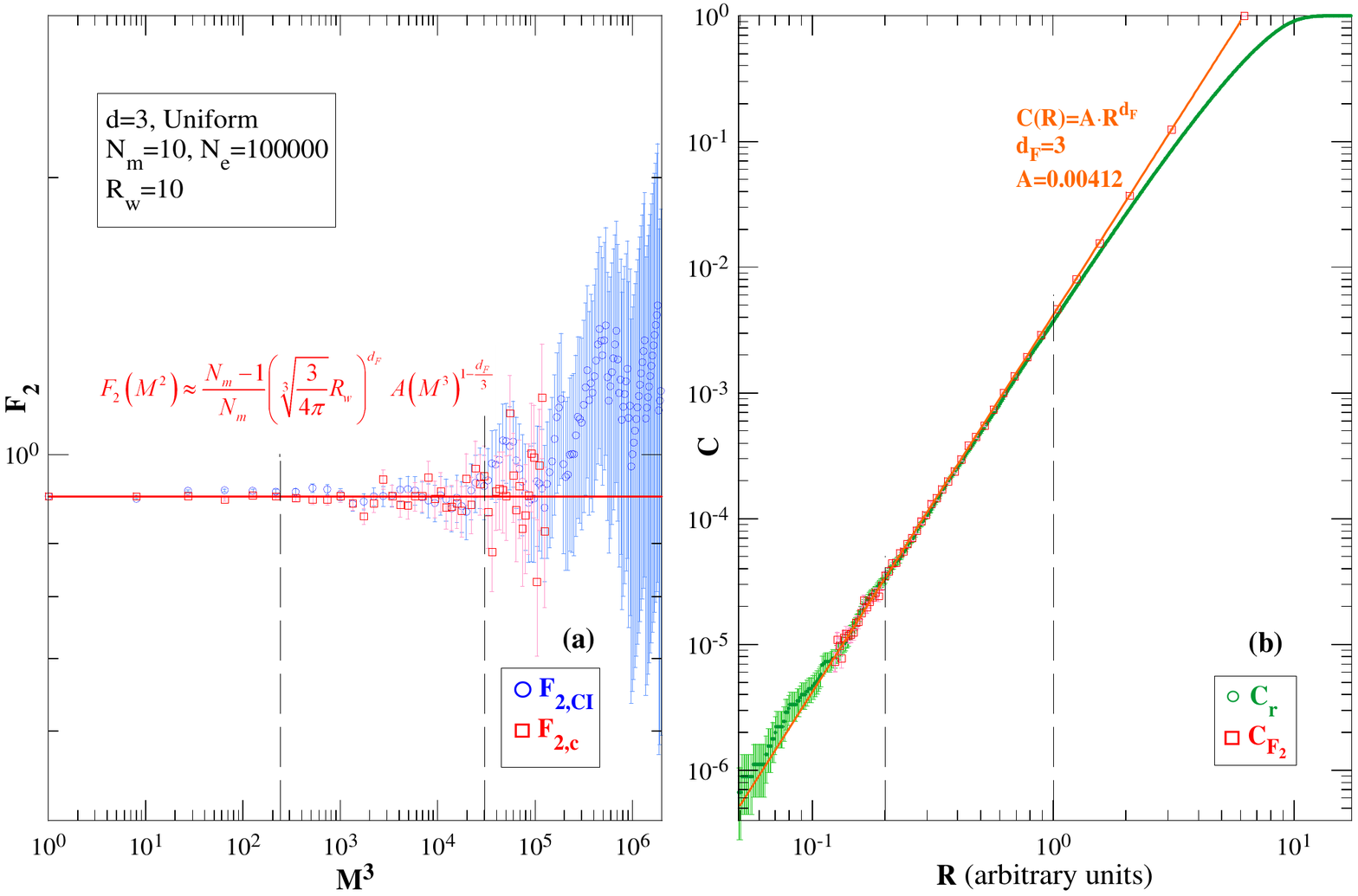}\\
\caption{\label{fig:3_uniform} {\small Analysis of a Uniform data set of $N_e=100000$ events and
multiplicity $N_m=10$ in an embedding space of $d=3$ dimensions.}}
\end{minipage}
\vspace{-0.0cm}
\end{figure}

\begin{figure}[H]
\begin{minipage}[]{1.\textwidth}
\centering
\includegraphics[scale=0.6,trim=2.cm 2.5cm 2.cm 0.cm,angle=0]{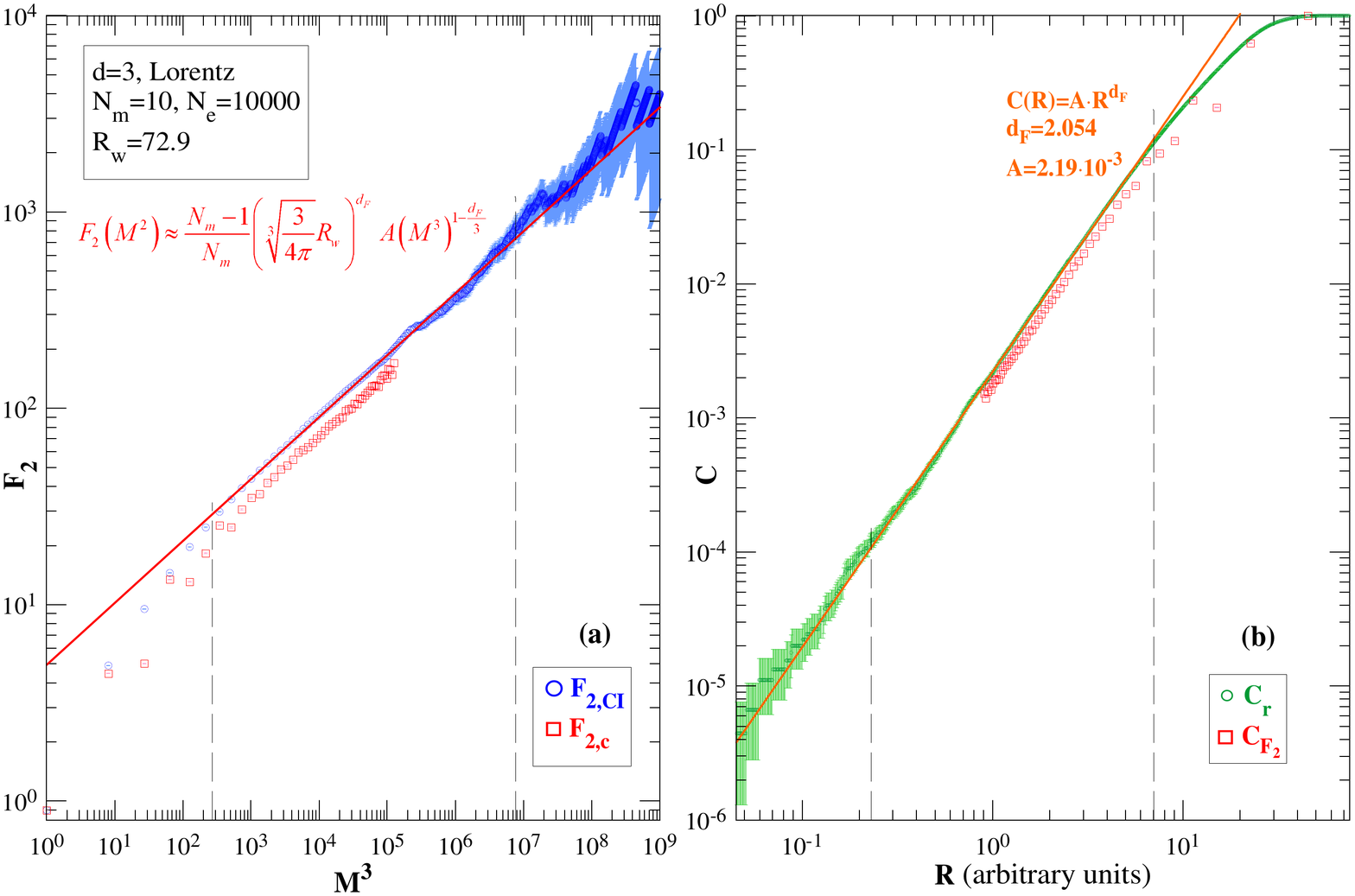}\\
\caption{\label{fig:3_lorentz} {\small Analysis of a Lorentz data set of $N_e=10000$ events and
multiplicity $N_m=10$ in an embedding space of $d=3$ dimensions.}}
\end{minipage}
\vspace{-4.cm}
\end{figure}

\clearpage


\vspace{-0.5cm}
\begin{table}[h]
\begin{minipage}[]{1.\textwidth}
\centering
\begin{tabular}{|c|c|c|c|c|c|} \hline
$d$ & data set & $d_{F,CI}$          & $d_{F,F_2}$         & $d_{F,th}$                  & Ref.            \\ \hline
1   & Uniform  & 0.99407$\pm$0.00019 & 1.00140$\pm$0.00012 & 1.072$\pm$0.037,1           & \cite{dimU1-H-L}\\ \hline
1   & L\'{e}vy & 0.3344$\pm$0.0004   & 0.34342$\pm$0.00006 & 0.33333                     &                 \\ \hline
2   & Henon    & 1.1974$\pm$0.0005   & 1.2174$\pm$0.0005   & 1.21$\pm$0.01               & \cite{CI}       \\
    &          &                     &                     & 1.220$\pm$0.036,1.258       & \cite{dimU1-H-L}\\
    &          &                     &                     & 1.21$\pm$0.01,1.25$\pm$0.02 & \cite{dimH-L}   \\ 
    &          &                     &                     & 1.21,1.22,1.28,1.52,1.36    & \cite{dimH-I-L} \\ \hline
2   & Ikeda    & 1.6749$\pm$0.0007   & 1.6677$\pm$0.0008   & 1.68,1.72,1.71              & \cite{dimH-I-L} \\ \hline
3   & Uniform  & 2.9448$\pm$0.0025   & 3.001$\pm$0.010     & 3                           &                 \\ \hline
3   & Lorentz  & 2.054$\pm$0.002     & 2.043$\pm$0.005     & 2.05$\pm$0.01               & \cite{CI}       \\ 
    &          &                     &                     & 2.049$\pm$0.096,2.062       & \cite{dimU1-H-L}\\ 
    &          &                     &                     & 2.05$\pm$0.01               & \cite{dimH-L}   \\ 
    &          &                     &                     & 2.04,$>$1.90,2.06,2.05,     & \cite{dimH-I-L} \\ 
    &          &                     &                     & $<$2.14,$<$2.12             & \cite{dimH-I-L} \\ \hline  
\end{tabular}
\vspace{-0.0cm}
\caption{\label{tab:d_F} {\small Results for the fractal dimension $d_F$ calculated by fits on the
data depicted on Figs.~\ref{fig:1_uniform}-\ref{fig:3_lorentz} between the slashed lines. 
$d_{F,CI}$ are extracted by fits on the $C_r$ data in graphs (b) and $d_{F,F_2}$ are extracted by fits
on the data $F_{2,CI}$ in graphs (a). For comparison estimations from corresponding references are provided.}}
\end{minipage}
\end{table}

\noindent
produce the peek. 
In other cases, like the Henon or the Ikeda
attractors, the one-particle distribution exhibit several sharp peeks 
at different locations. The grid at certain partitions may or may not
split these sharp peeks, producing, in the absence of grid averaging,
notable differences in adjacent partitions at low $M$.

\subsection{Limitations}\label{subsection:limit}

An infinitely large data set can enable us to access the attributes of the set even at
infinitesimal scales. However, we have at hand or can produce a finite number of data points.
Inevitably, this leads us to a situation where, as we progress our analysis towards lower
scales, the statistical fluctuations will become higher. Then, at even lower scales we will find zero
points to fall into our bins. This is the ``zero bin'' effect. We can make an estimation of the
scale where this effect takes place. The constant $A$, in the correlation integral approximation of
eq.~(\ref{eq:C-A}), is related to a scale $R_{max}$, which is connected to the size of the subspace
containing the whole data set
to be analysed, since at that scale we must have $C(R_{max})=1$. So we can set 
$A \simeq R_{\max}^{-d_F}$ and $C$ can be written as
\begin{equation}{\label{eq:C(R)-Rmax}}
C(R) \simeq \left(\frac{R}{R_{\max}} \right)^{d_F}\;.
\end{equation}
Then, let $R_{\min}$ be the scale which corresponds to a bin where we will find, on the average, 
one pair if we search the whole 
data set. Beyond that scale we would not expect to find a pair, so the most probable value for $C$ 
would be zero. Then, using eq.~(\ref{eq:C(R)-Rmax}) and the definition of $C$, eq.~(\ref{eq:CI}), we get
\[
C(R) = \frac{2}{{{{\left\langle {{N_m}\left( {{N_m} - 1} \right)} \right\rangle }_e}}}{\left\langle {{N_p}(R)} 
\right\rangle _e} \Rightarrow
C(R) = \frac{2}{{{{\left\langle {{N_m}\left( {{N_m} - 1} \right)} \right\rangle }_e}}}\frac{{\sum\limits_e {{N_p}(R)} }}
{{{N_e}}} \Rightarrow 
\]
\begin{equation}{\label{eq:empty_bin}}
\Rightarrow \left(\frac{R_{\min}}{R_{\max}} \right)^{d_F}
 \simeq \frac{2}{{{{\left\langle {{N_m}\left( {{N_m} - 1} \right)} \right\rangle }_e}}}\frac{1}{{{N_e}}} \Rightarrow
R_{\min} = R_{\max}
\left[ \frac{2}{{{N_e}\left\langle N_m\left(N_m-1 \right)\right\rangle_e}} \right]^{\frac{1}{d_F}}\;.
\end{equation}

Eq.~(\ref{eq:empty_bin}) reveals that the increase of number of events $N_e$ and the number of 
multiplicities
$N_m$ can help us access lower scales. However, the fractal dimension of the set $d_F$ is crucial,
since when it acquires higher values, the data set is depleted more rapidly over the scales.
Also, we can see this in Figs.~\ref{fig:1_uniform}-\ref{fig:3_lorentz}. For example, comparing
Figs.~\ref{fig:1_uniform} and \ref{fig:3_uniform}, which both describe uniform distributions
with the same number of data points, we see that the ratio $\frac{R_{\max}}{R_{\min}}$ acquires a much
lower value in the 3-dimensional space compared to the 1-dimensional one, so the statistical fluctuations
become prominent at higher scales in three dimensions.

Our aim is to always be able to reach the ``zero bin'' limit in the data analysis with the techniques 
presented in this paper, so that there will be no information left uncovered.

\subsection{Efficiency Considerations}\label{subsection:effic}

During the data analysis in experiments it is needed to make corrections in the final
results for the detectors efficiency, i.e. for tracks which have not been measured.
This amounts to finding out how the results are altered if the number of multiplicities 
$N_m $ per event is
increased by a certain ratio $z>1$,
so $(z-1)N_m$ represents the number of
missed tracks per event.
We will discuss how the $F_2$ and $C$ curves are affected in such a case. We will consider
first the case where the number of multiplicities is fixed, i.e. all the events contain
the same number of tracks. We will denote by unprimed quantities the initial ones before the
increase in their values and by primed quantities the increased ones.
The multiplicities $N_m^{\prime}=z N_m$ correspond to an increase
to the total number of pairs per event by a factor 
$z^{\prime}=z^2 \frac{N_m-1/z}{N_m-1}$, since
$N_p^{\prime}=\frac{1}{2}N_m^{\prime}(N_m^{\prime}-1)=
\frac{1}{2}z^2 N_m(N_m-1/z)=
z^2 \frac{(N_m-1/z)}{N_m-1}\frac{1}{2}N_m(N_m-1)=z^{\prime} N_p$. 
There
are three possibilities:
\newline (a) The number of pairs per event at scale $R$, $N_p(R)$, for all scales, is increased 
by the same ratio 
and at the same time all the additional multiplicities do not produce a greater distance
between them compared to the previous ones. This means that the additional pairs will 
be distributed among the 
initial set of scales and proportionally to the initial number of pairs per scale. Thus, 
all $N_p(R)$ are increased by the same ratio $z^{\prime}$. Then from eq.~(\ref{eq:CI}) 
it follows that
\begin{equation}{\label{eq:C_eff1}}
C^{\prime}(R)=\frac{2}{N_m^{\prime}(N_m^{\prime}-1)} \left\langle N_p^{\prime}(R)\right\rangle_e =
\frac{2 z^{\prime}}{z^{\prime} N_m(N_m-1)}  
\left\langle N_p(R)\right\rangle_e=
C(R) \Rightarrow C^{\prime}(R)=C(R)\;.
\end{equation}
Thus, the $C(R)$ curve will remain completely unchanged.
\newline (b) The number of pairs per event at scale $R$, for all scales, is increased 
by the same ratio 
but the existence of the additional multiplicities produce some greater distance
between them compared to the previous ones. This means that the additional pairs will 
be distributed among a greater number 
of scales and proportionally to the initial number of pairs per scale. The $C^{\prime}(R)$
curve will reach the maximum unit value at a greater scale. Now, 
the number of pairs at scale $R$, $N_p^{\prime}(R)$ will be increased by a constant ratio 
$z^{\prime\prime}<z^{\prime}$ and we will have
\begin{equation}{\label{eq:C_eff2}}
C^{\prime}(R)=\frac{2}{N_m^{\prime}(N_m^{\prime}-1)} \left\langle N_p^{\prime}(R)\right\rangle_e
=\frac{2 z^{\prime\prime}}{z^{\prime} N_m(N_m-1)}  \left\langle N_p(R)\right\rangle_e=
\frac{z^{\prime\prime}}{z^{\prime}}C(R)
\Rightarrow C^{\prime}(R)<C(R)
\;.
\end{equation}
The $C(R)$ curve retains the same shape but it is shifted to the right (to greater scales).
\newline (c) The number of pairs per event do not increase proportionally to the
initial number of pairs and greater maximum scales may be introduced to the system\footnote{A simulated system produced by
a series of few successive L\'{e}vy steps with
the probability distribution (\ref{eq:Levy}) is an example where the $C$
curve will depend on the number of steps which form an event.}.
In this case the shape of $C$ will change with the increase of additional 
multiplicities depending on the particular system. 

In all the above cases, the $F_2$ curve will
change according to eq.~(\ref{eq:F2-C}).
In (a) the shape of $F_2$ will remain the same, but it will 
increase by a factor 
$\frac{a^{\prime}_m}{a_m}=\frac{N_m-1/z}{N_m-1}$.
In (b) the shape of $F_2$ will remain the same. Its value will 
change by a factor $\frac{z^{\prime\prime}}{z^2}$. This will
lead to an increase, to a decrease or will leave the curve
unchanged according to the value of terms $z^{\prime\prime}$ and
$z^2$.
In (c) the shape of $F_2$ will in general change, according to the 
change of the shape of the $C$ curve and eq.~(\ref{eq:F2-C}).

To see the effect of increasing the multiplicity in the case
where the number of tracks per event is not fixed, one has to
divide the events in subsets of events with fixed multiplicity.
Then, the increase of the multiplicity by a certain ratio should be 
applied at each subset and explore its effect.

We investigate the above considerations by application to specific
systems. Our results are shown in Figures \ref{fig:effic_uni} and
\ref{fig:effic_hen}, where the $F_2$ results are depicted in
graphs (a) and the $C$ results in graphs (b). The error depiction is 
dropped in order to show clearly the important information. \linebreak

\vspace{-0.2cm}
\begin{figure}[H]
\begin{minipage}[]{1.\textwidth}
\centering
\includegraphics[scale=0.75,trim=3.2cm 9.4cm 3.2cm 0.cm,angle=0]{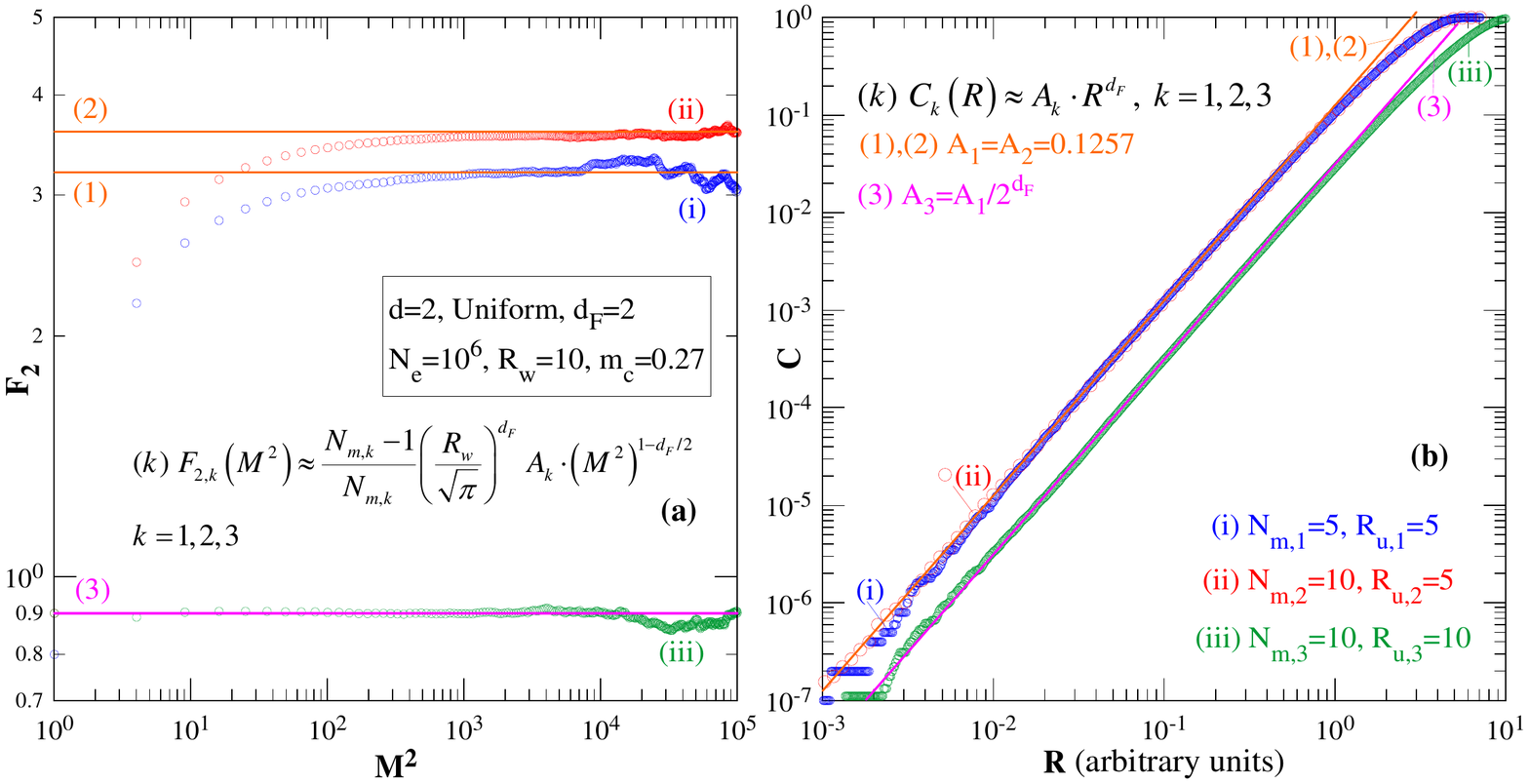}\\
\caption{\label{fig:effic_uni} {\small Analysis of Uniform data sets of $N_e=10^6$ 
events. Case (i) has points of
multiplicity $N_{m,1}=5$ distributed in a rectangle of side $R_{u,1}=5$. 
In similar manner we have $N_{m,2}=10$ and $R_{u,2}=5$ for case (ii) and
$N_{m,3}=10$ and $R_{u,2}=10$ for case (iii).}}
\end{minipage}
\vspace{-0.3cm}
\end{figure}

\begin{figure}[H]
\begin{minipage}[]{1.\textwidth}
\centering
\includegraphics[scale=0.6,trim=1.0cm 3.6cm 0.0cm 2.9cm,angle=0]{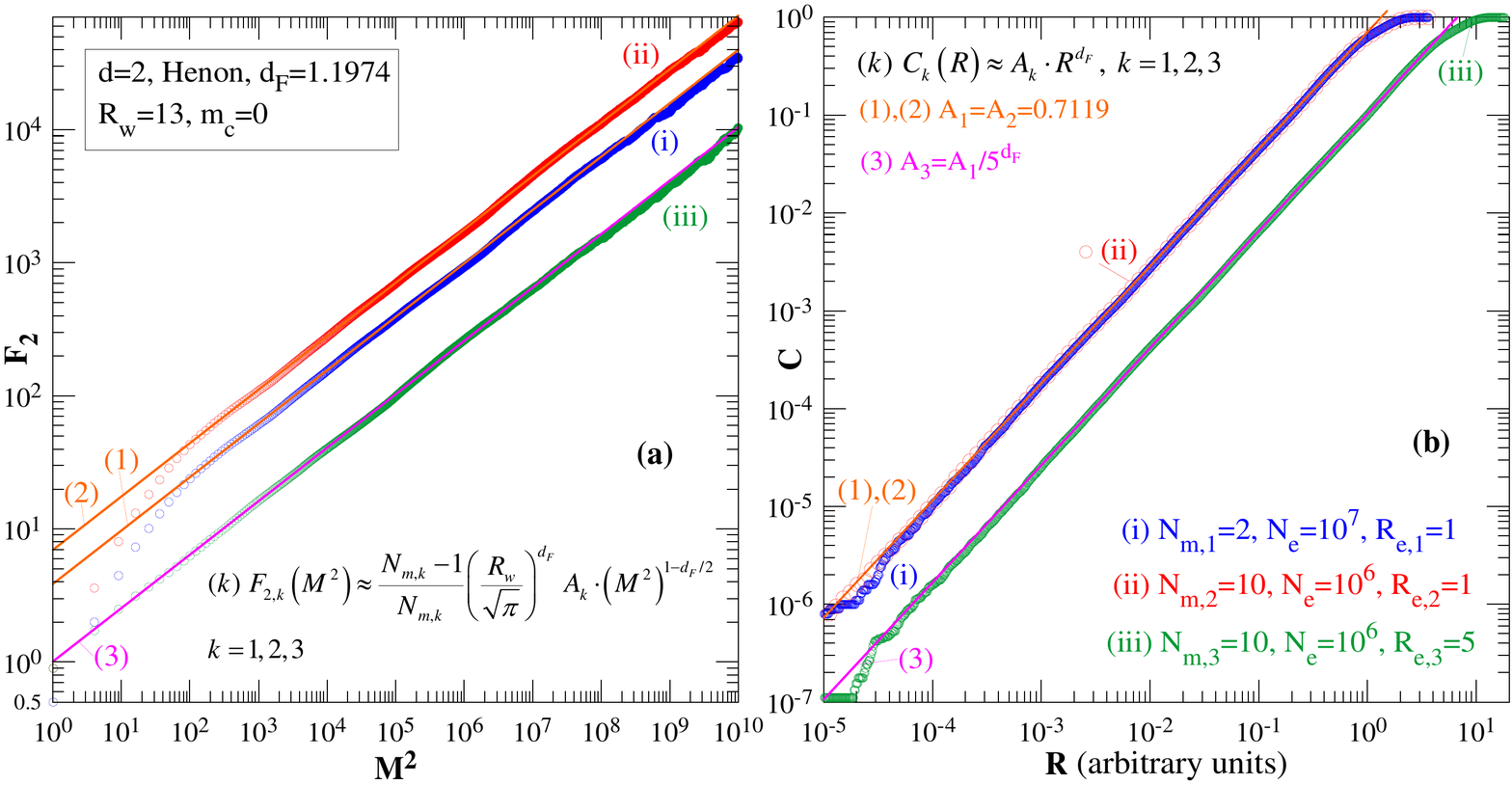}\\
\caption{\label{fig:effic_hen} {\small Analysis of Henon data sets. Case (i) 
involves $N_e=10^7$ events with multiplicity $N_{m,1}=2$. 
In similar manner case (ii) corresponds to $N_e=10^6$ and $N_{m,2}=10$.
Case (iii) is similar to case (ii) but the points are expanded in space by a
factor $R_{e,3}=5$.}}
\end{minipage}
\end{figure}

\noindent
With latin numbers we represent the curves produced from our
simulation data and with arabic numbers we represent the
theoretical curves that approximate our data curves away from
the boundaries (lower scales or greater partitions).

Firstly we deal with a uniform system in a 2-dimensional 
rectangle and  the relevant results are shown in Figure \ref{fig:effic_uni}.
We retain for all $F_2$ cases the same analysis window
$[-5,5]\times[-5,5]$, with $R_w=10$.
Initially, we produce a set of points in the rectangle
$[-2.5,2.5]\times[-2.5,2.5]$, i.e. with side $R_{u,1}=5$ 
and multiplicity $N_{m,1}=5$. The corresponding $C$ and $F_2$
curves are shown as curves (i) in graph 
(b) and (a), respectively. Then, we increase the multiplicity per
event to $N_{m,2}=10$, still confined to the same rectangle with
side $R_{u,2}=5$. The corresponding $C$ curve, (ii) in
graph (b) remains unchanged with respect to curve (i) in the same
graph, while the $F_2$ curve, (ii) in graph (a), is increased by a
factor $\frac{a^{\prime}_m}{a_m}=\frac{4.5}{4}$. 
In the third case we produce uniformly
distributed points in the rectangle
$[-5,5]\times[-5,5]$, i.e. with side $R_{u,3}=10$. We, now,
observe that the $C$ curve, (iii) in graph (b), while it
retains the same slope, which corresponds to $d_F=2$, is
shifted to greater scales. In this situation we can evaluate
exactly the magnitude of the shift. Curve (i) is approximated by 
the relation $C_1(R)\simeq A_1 R^2$. The expansion of the uniform
2-d rectangle by a factor of 2 leads the $C$ curve of case (iii) to 
acquire the maximum unit value at 2 times the initial value. Thus,
Curve (iii) is approximated by the relation $C_3(R)\simeq A_3 R^2=
\frac{A_1}{2^2} R^2$. The $F_2$ curve (iii) compared to curve (i)
is increased by a factor $\frac{a^{\prime}_m}{a_m}=\frac{4.5}{4}$ due to
the multiplicity increase and decreased by a factor $\frac{A_1}{A_3}=\frac{1}{4}$
due to the change of the $C$ curve. The overall effect is a total decrease by a 
factor of $\frac{4.5}{16}$.

Secondly we take points from the Henon attractor. The relevant results are shown in Figure \ref{fig:effic_hen}.
We retain for all $F_2$ cases the same analysis window
$[-6.5,6.5]\times[-6.5,6.5]$, with $R_w=13$.
Initially, we produce $N_{m,1}=2$ points per event using the Henon equations.
The result is that all points always fall inside an area in the ($x-y$) plane
which can be enclosed by the rectangle $[-1.3,1.3]\times[-1.3,1.3]$.
The corresponding $C$ and $F_2$
curves are shown as curves (i) in graph 
(b) and (a), respectively. Then, we increase the multiplicity per
event to $N_{m,2}=10$. The resulting points are still enclosed in the same area. 
The corresponding $C$ curve, (ii) in
graph (b) remains unchanged with respect to curve (i) in the same
graph, while the $F_2$ curve, (ii) in graph (a), is increased by a
factor $\frac{a^{\prime}_m}{a_m}=\frac{1.8}{1}$. 
In the third case we produce again $N_{m,3}=10$ Henon points per event, but, now,
we multiply the $x$ and $y$ coordinates of each point with a factor
$R_{e,3}=5$. The result is that the points retain the fractal dimension of the
Henon  
set, but they are enclosed to an area expanded by a factor 5,
compared to
case (i). Thus, the $C$ curve, (iii) in graph (b), 
retains the same slope, which corresponds to $d_F\simeq1.2$, but it is
shifted to greater scales. To evaluate the magnitude of 
the shift, we observe
that curve (i) is approximated by the
relation $C_1(R)\simeq A_1 R^{d_F}$, while curve (iii) is approximated by the 
relation $C_3(R)\simeq A_3 R^2=
\frac{A_1}{5^{d_F}} R^{d_F}$. The $F_2$ curve (iii) compared to curve (i)
is overally decreased.

\section{A powerful ``microscope''}\label{sec:micro}

A lot of times in correlation phenomena we want to detect as signal the fractal structure of
a data set, exhibited by an exponent $d_F$, the fractal dimension of the set.
Usually, we have at hand experimental data which do not only contain pure data of the
wanted signal, but, also, data behaving as ``noise'' for our purpose and which may exist
 at a much
higher percentage. Let this noise data be described by a slope of the $C$-curve in logarithmic plot equal to $d_n$. 
If this noise is attributed to uncorrelated random processes defined in $d$ dimensions, then $d_n=d$,
where $d$ is the dimension of the 
embedding space. In that case, since the fractal dimension is always lower than the embedding space dimension,
$d_F<d_n$. Then, the correlation integral of our data set can be 
approximated by a relation of the form
\begin{equation}\label{eq:C-AB}
C(R) \approx C_{cr}(R)+C_n(R) \equiv A\cdot R^{d_F} + B\cdot R^{d_n}\;,
\end{equation}
where $C_{cr}(R) = A\cdot R^{d_F}$ is the part which contains our signal, called from now on as critical part
and $C_n (R) = B\cdot R^{d_n}$ 
is the noise part. 
The two exponents $d_F$ and $d_n$ determine the behaviour of $C(R)$ at the scales $R$.
As scale $R$ increases, the effect of $d_n$ increases and the effect of $d_F$ weakens.
The opposite is true as scale $R$ decreases.
Indeed, we can observe this by defining the ratios
\begin{equation}\label{eq:e_cr}
e_{cr}(R)=\frac{N_{cr.\;pairs}(R)}{N_{all\;pairs}(R)}=\frac{C_{cr}(R)}{C(R)}=
\frac{A\cdot R^{d_F}}{A\cdot R^{d_F} + B\cdot R^{d_n}}
\end{equation}
\begin{equation}\label{eq:e_n}
e_{n}(R)=\frac{N_{n.\;pairs}(R)}{N_{all\;pairs}(R)}=\frac{C_{n}(R)}{C(R)}=
\frac{B\cdot R^{d_n}}{A\cdot R^{d_F} + B\cdot R^{d_n}}\;,
\end{equation}
where $e_{cr}(R)$ and $e_{n}(R)$ is the ratio of the critical and noise pairs to the total pairs, respectively,
 at a specific scale $R$.
Differentiating with respect to the scale $R$ we get
\begin{equation}\label{eq:de_cr}
\frac{d[e_{cr}(R)]}{dR}=
\frac{A\cdot B \cdot R^{d_F+d_n -1}}{\left(A\cdot R^{d_F} + B\cdot R^{d_n}\right)^2} (d_F - d_n)<0
\end{equation}
\begin{equation}\label{eq:de_n}
\frac{d[e_{n}(R)]}{dR}=
\frac{A\cdot B \cdot R^{d_F+d_n -1}}{\left(A\cdot R^{d_F} + B\cdot R^{d_n}\right)^2} (d_n - d_F)>0
\end{equation}
Thus, $e_{cr}(R)$ is a descending function and $e_{n}(R)$ is an 
ascending function with respect to the scale $R$. 
Also, $0<e_{cr}(R)<1$, $0<e_{n}(R)<1$, $e_{cr}(R)+e_{n}(R)=1$ and 
\begin{equation}\label{eq:e_cr_lim}
\underset{R \to ^{\scriptstyle \;0}_{\scriptstyle \infty}}{\lim} e_{cr}(R)=
\underset{R \to ^{\scriptstyle \;0}_{\scriptstyle \infty}}{\lim} 
\left(1 + \frac{B}{A}\cdot R^{d_n-d_F}\right)^{-1}
=\;^{\textstyle 1}_{\textstyle 0}
\end{equation}
\begin{equation}\label{eq:e_n_lim}
\underset{R \to ^{\scriptstyle \;0}_{\scriptstyle \infty}}{\lim} e_{n}(R)=
\underset{R \to ^{\scriptstyle \;0}_{\scriptstyle \infty}}{\lim} 
\left(1 + \frac{A}{B}\cdot R^{d_F-d_n}\right)^{-1}=
\underset{R \to ^{\scriptstyle \;0}_{\scriptstyle \infty}}{\lim} 
\left[1 + \frac{A}{B}\cdot \left(\frac{1}{R}\right)^{d_n-d_F}\right]^{-1}
=\;^{\textstyle 0}_{\textstyle 1}
\end{equation}
So, at infinite scales our data set behaves as a purely noise set and at infinitesimal scales as
a purely critical one. We can use $e_{cr}(R)$ as an estimator of the approach to
the fractal behaviour of our data, starting from high scales and moving towards low ones.
Moving in the same direction, $e_{n}(R)$ can be used as an estimator of the weakening of the noise
behaviour of our data. Indeed, the slope $\kappa(R)$ of $C$ at scale $R$, in a logarithmic graph, can be
expressed with respect to these estimators as
\begin{equation}\label{eq:C_slope}
\kappa(R)=\frac{d\ln(C)}{d\ln(R)}=d_F e_{cr}(R)+d_n e_{n}(R)=
d_F e_{cr}(R) + d_n \left[ 1-e_{cr}(R) \right]\;.
\end{equation}
Thus, moving in the aforementioned direction, $e_{cr}(R)$ describes how the $C$-slope, an easily
observed quantity, changes from $d_n$ at high scales to $d_F$ at low ones.

We may want to determine the specific scale $R_q$ at which $e_{cr}$ acquires a certain value $q$. So
\begin{equation}\label{eq:e_cr_q}
e_{cr}(R_q)=q \Rightarrow \frac{A\cdot R_q^{d_F}}{A\cdot R_q^{d_F} + B\cdot R_q^{d_n}} =q \Rightarrow
A(1-q) R_q^{d_F}=BqR_q^{d_n} \Rightarrow
R_q = \left[\frac{A}{B} \frac{(1-q)}{q} \right]^{\frac{1}{d_n-d_F}}\;.
\end{equation}
Of special interest is the value $q=0.5$. At the relevant scale, $R_t \equiv R_{0.5}$, the data set 
transcends from one
behaviour to the other. For scales $R<R_t$ the critical behaviour becomes the dominant one.
This characteristic scale is
\begin{equation}\label{eq:e_cr_0.5}
R_t = \left(\frac{A}{B} \right)^{\frac{1}{d_n-d_F}}\;.
\end{equation}
Since $R_t$ may not always be reachable from the experimental data set, one may search for a scale, 
$R_s$, where the deviation from 
the purely ``noisy'' system starts to show. We can estimate this scale to be located at the point 
where this 
deviation exceeds the magnitude of the experimental errors. If the relative experimental error in
our measurements is $r$, we can set
\begin{equation}\label{eq:e_cr_r}
R_s \simeq R_r\;.
\end{equation}
A common experimental error of the order of 10\% would lead to:
\begin{equation}\label{eq:e_cr_0.1}
R_s = \left(\frac{9A}{B} \right)^{\frac{1}{d_n-d_F}}\;.
\end{equation}

With the use of eqs.~(\ref{eq:F2-C}) and (\ref{eq:C-AB}) we can approximate $F_2$ as
\begin{equation}\label{eq:F2-AB}
F_2(M^d) \simeq F_{2,cr}(M^d)+F_{2,n}(M^d) \equiv 
a_m \left(\beta_d R_w\right)^{d_F} A \left( M^d \right)^{1-\frac{d_F}{d}} +
a_m \left(\beta_d R_w\right)^{d_n} B \left( M^d \right)^{1-\frac{d_n}{d}}\;,
\end{equation}
where $F_{2,cr}$ and $F_{2,n}$ is the first and second term of $F_2$ which contains the critical
and noise contribution, respectively. If $d_n \simeq d$, then the noise part of $F_2$ will be
independent of $M^d$.

\begin{figure*}[h]
\vspace{-0.0cm}
\centering
\includegraphics[scale=0.58,trim=0.cm 3.7cm 0.cm 3.5cm,angle=0]{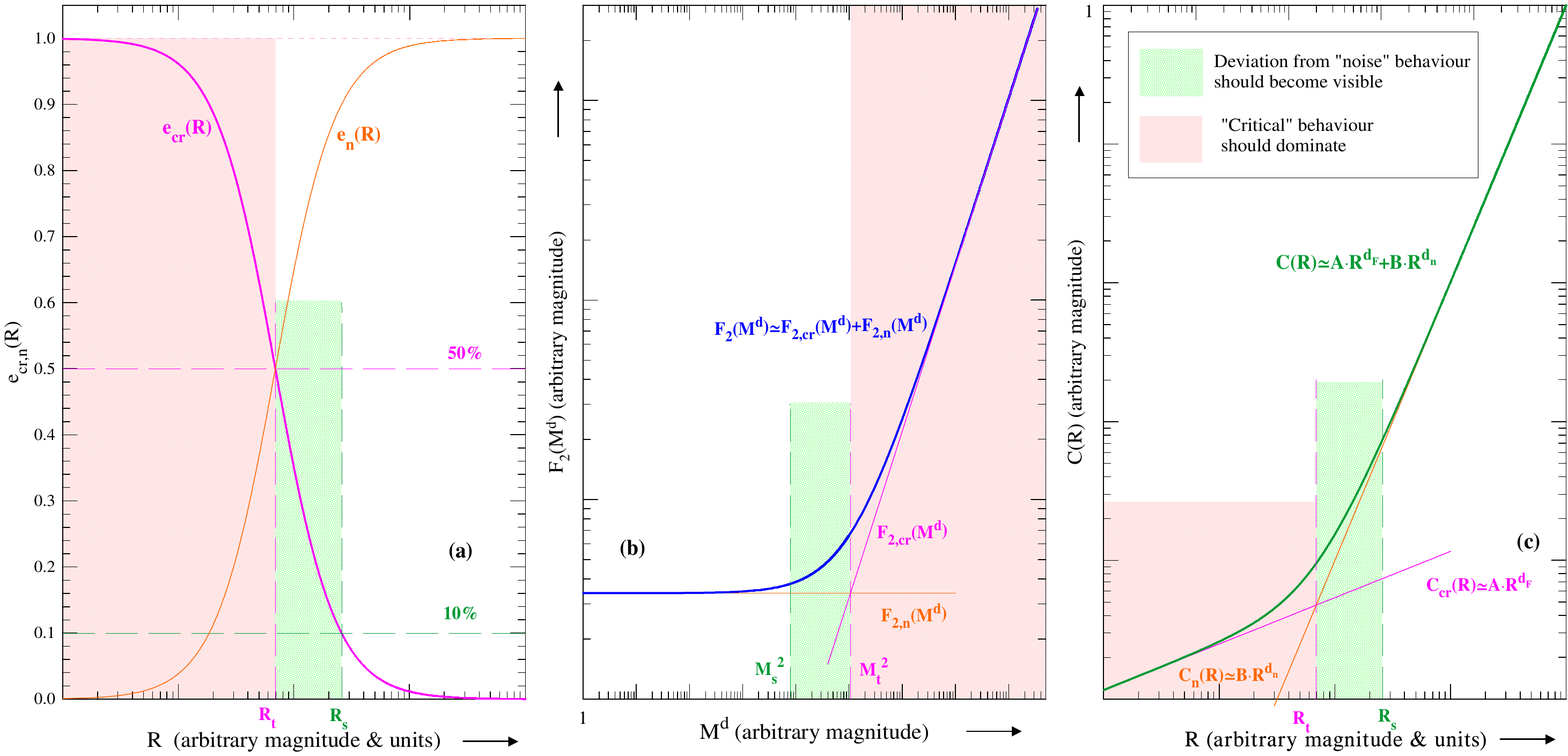}\\
\caption{\label{fig:ecr,F2,C_th} {\small Qualitative description of the behaviour of the estimators 
$e_{cr}(R)$, $e_{n}(R)$ (a), the factorial moments $F_2$ (b) and the correlation integral $C$ (c),
in a case where a system with high percentage of noise is analysed.}}
\end{figure*}

It may happen that our data set is such that at the higher scanned scales the system exhibits behaviour 
similar to a purely noise data set. A data set with a high percentage of noise increases the 
possibility that this will be the case.
Then, the only option we have to detect a critical behaviour is to direct our analysis towards the low
scales. A qualitative description of what we may encounter 
is exhibited in Fig.~\ref{fig:ecr,F2,C_th}.
The graphs of $C$ and $F_2$ are displayed in (c) and (b), respectively. $C$ at high scales
and $F_2$ at low partitions cannot be distinguished from the system of pure noise (i.e.~the 
slopes
of $C$ and $F_2$ are $d_n$ and $1-d_n/d$, respectively). As we
progress our analysis to lower scales (or higher partitions), we arrive at a scale $R_s$ (or
partition $M_s^d$)
where the deviation from the purely noise system should become apparent. 
This is shown by the begging of
change of slope towards lower values in $C$ and higher values in $F_2$.
Progressing further, we pass through a scale $R_t$ (or partition $M_t^d$), entering at a
territory where the critical set starts to dominate and the slopes of $C$ and $F_2$ and tend towards
$d_F$ and $1-d_F/d$, respectively.
It is realised that we have to exhaust our analysis down to the lower scales to extract all possible
information hidden in the data.
Of course, the lower scale analysis will be stopped, either by the approach to the ``empty bin'' limit
(eq.~(\ref{eq:empty_bin})), where the statistical fluctuations will increase immensely, or by the 
approach to scales corresponding to the detector resolution.
In Fig.~\ref{fig:ecr,F2,C_th}(a) we depict the estimators $e_{cr}(R)$ (eq.~(\ref{eq:e_cr})) and
$e_{n}(R)$ (eq.~(\ref{eq:e_n})). The first can be used as an estimator of the approach to the
region of scales with the critical behaviour. In such a scenario, as this depicted in 
Fig.~\ref{fig:ecr,F2,C_th}, the value of $B$ can be determined by a fit on the $C$-curve
at high scales,
while the evaluation of $A$ necessitates the appearance of a segment of the $C$-curve clearly
departing from the territory with slope $d_n$.

We apply the above to a more concrete example. We form by simulation a data set of $N_e=10^6$
events with multiplicity $N_m=26$, consisting mostly of noise events and with some events containing
the signal to be detected (critical events)
in an embedding space of $d=2$ dimensions.
A critical event is formed by $N_m$=26 L\'{e}vy walks in two 
1-dimensional spaces, in the same 
way as it was done in section \ref{sec:app} (see eq.~(\ref{eq:Levy})) and using the parameters
$\nu=1/6$, $b=10^{-7}$ and $x_c=10^{-1}$.
The final coordinates of our points consist of one coordinate from one 1-dimensional space and
one from the other. This external product is expected to form a fractal set with dimension
$d_f=\frac{1}{6}+\frac{1}{6}=\frac{1}{3}$. We produce 150000 such critical events. A
noise event is produced by $N_m=26$ uniform points in the 2-dimensional rectangle 
$[-0.9,0.9] \times [-0.9,0.9]$. 
To form the data set which contains the signal combined with noise, we pick randomly 
10 events from the 
critical ones, while the rest events are uniform ones, so that to have the total number of $10^6$.
To form the data set which contains only noise, we separately produce $10^6$ uniform events.
Our results of the analysis of these data sets are depicted in Fig.~\ref{fig:F2_C_sim}.
On the graphs the quantities with an argument in parenthesis ($(R)$ in graph (b) and $(M^2)$ in (a))
depict curves which are theoretical approximations on the data, according to eqs.~(\ref{eq:C-AB})
and (\ref{eq:F2-AB}).
The quantities without arguments depict the actual calculations on the data.
These calculations for the correlation integral $C$ were carried out with the ring technique,
involved $N_D=$3500 estimation points and were completed with error calculation in 129 s.
The calculations for the factorial moments $F_2$ were carried out with the ring technique
and the correspondence to $C$\footnote{Here we have set in eq.~(\ref{eq:M-m_f})
$m_f=0.25$.}, involved $N_D=$100000 estimation points and were completed with error 
calculation in 2218 s. The corresponding calculations for $C$ and $F_2$, without the error calculation,
were carried out within 62 and 39 s, respectively\footnote{Recorded times in this section are all measured in the
same computing machine. They correspond to the calculation of the data set with signal and noise 
and are expected to be equal to the corresponding times for the data set of pure noise, since this data 
set is similar to the previous one with respect to all involved parameters.}.

On graph (b) we mark the scale $R_s\simeq5.7\cdot10^{-3}$, where we expect to see indications for
departure from the noise 
behaviour and $R_t\simeq1.5\cdot10^{-3}$, where we enter the domain of
the critical behaviour. The partitions for $F_2$ are $M_s\simeq179$ and $M_b\simeq670$,
respectively.
On the upper left of graph (a) we depict $\Delta F_2\equiv F_2-F_{2,n}$, which is estimated from 
the $F_2$ of the data 
containing signal and noise with subtraction of $F_{2,n}$, which corresponds to the pure noise data.
As expected, $\Delta F_2$ rises proportionally to $\left(M^d\right)^{1-\frac{d_F}{d}}$.
We perform estimation of this quantity up to the partitions where we have pure noise data
(the exponent $d_n$, which is higher than $d_F$, causes the empty bin effect to appear at the pure noise data
at lower partitions than the data which contain signal and noise).
We observe a gradually strengthened 
signal at higher values of $M^d$, recorded at the increasing value of $\Delta F_2$.
It should be noted that in analysis of real data, in which the statistics may be much less than in the simulation,
the increased statistical fluctuations will hinder $\Delta F_2$ to appear clearly at low $M$ partitions, due to
its low magnitude. The analysis has to proceed to larger $M$, so that the magnitude of $\Delta F_2$ will
increase beyond the magnitude of statistical errors and so the trend of the curve will appear clearly.

\begin{figure*}[h]
\centering
\includegraphics[scale=0.65,trim=1.6cm 1.2cm 1.3cm 0.cm,angle=0]{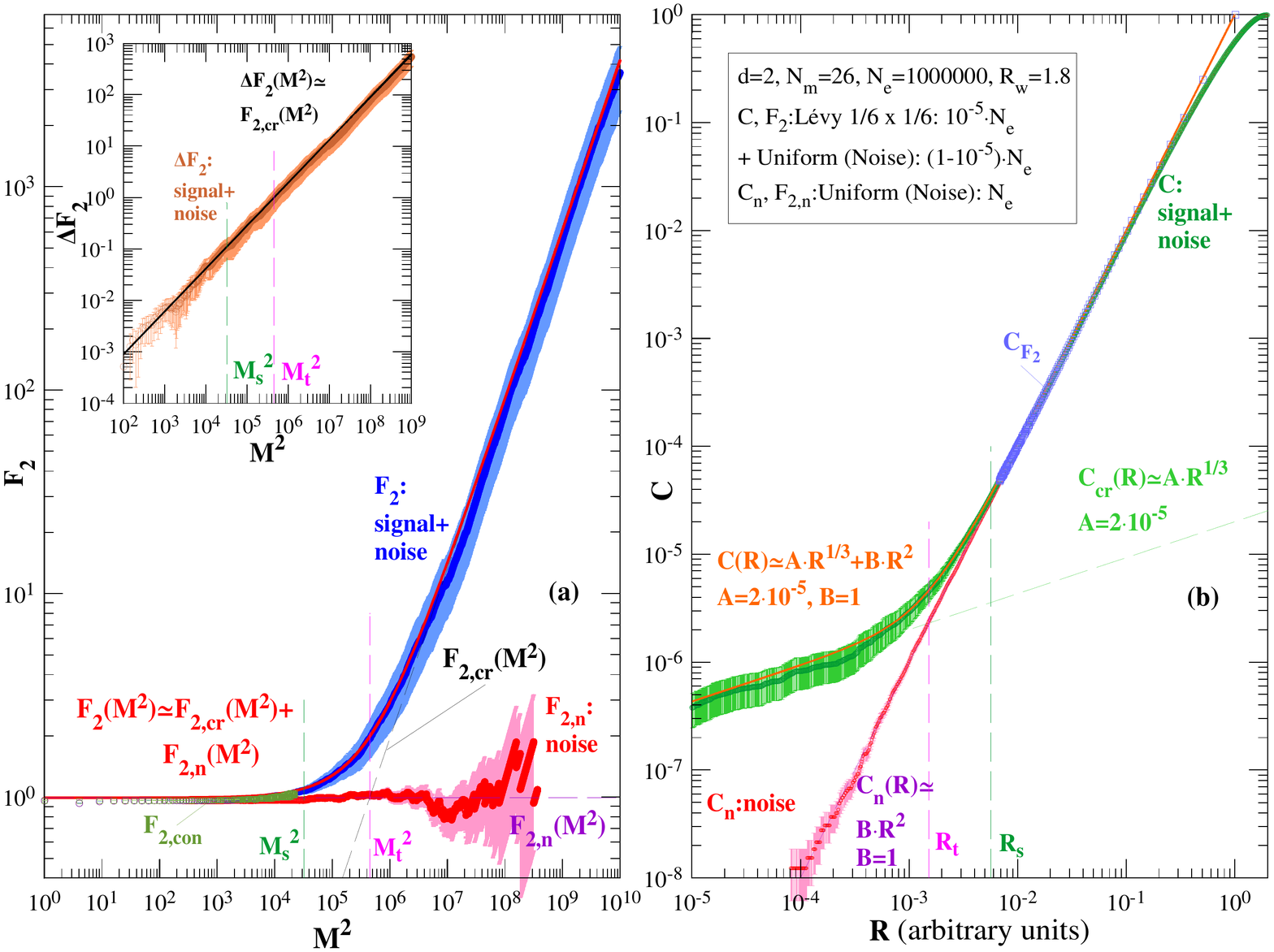}\\
\caption{\label{fig:F2_C_sim} {\small Analysis of a data set from a simulation in an embedding space
of $d=2$ dimensions, 
which contains critical L\'{e}vy events of $d_F=1/3$ and a high percentage of noise, which is a uniform 
distribution with $d_n=2$. The factorial moments $F_2$ are depicted in (a) and the correlation 
integral $C$ in (b).}}
\end{figure*}

On graph (a) we, also, depict the calculations of $F_2$ for the signal and noise data set with the 
conventional way, up to partition $M=150$, (marked as $F_{2,{\rm con}}$) with the corresponding 
mapping to $C$
in graph (b) (marked as $C_{F_2}$). The calculation, performed with no grid averaging which
would increase time consumption, took about 15247 s to complete. 
However, up to this partition we do not observe a clear sign of critical behaviour.
We estimate by extrapolation that in order to reach calculations with the
conventional routine up to $M_s\simeq179$
it would take $\sim$7 hours, up to $M_b\simeq670$ $\sim$15 days, while to reach 
$M=100000$, which we easily achieve with our technique, it would require time of the order of magnitude 
of tens of thousands years.

Thus, we see that, with the ring technique and the $F2-C$ correspondence, we can address
large amount of data and scan them up to the very low scales or very high partitions.
We have at our hand a very effective ``microscope'', which enables us to extract all the information
hidden in our data, down to the accuracy set by our measuring equipment.
This attribute is due to the negligible time consumption compared to the conventional techniques.

\section{Conclusions} \label{sec:conclu}

We have developed a general mapping between correlation integral and second order scaled
factorial moments which constitute two important tools used for the description of correlations
in arbitrary data sets.
This relation enables us to pass the results of the estimation of one quantity to the other.
We are, also, able to form approximations of $C$ and $F_2$ that describe the data.

In this way, we can evaluate $F_2$ essentially by performing calculations similar to the correlation integral $C$,
thus using moving disks, which monitor the data location, instead of a fixed grid.
In this way, we can avoid uncertainties due to the grid location and avoid missing pairs corresponding 
to certain scales or partitions.

Further, we notice that the correlation integral $C$ provides a clear rating of the distance of pairs
to scales. Inspired by this, we develop a new computational technique, 
which counts first the pairs within rings of width $dr$ around $r$ (with $0<r<R$) and then 
builds the results for disks of radius $R$.
The advantage is an extreme reduction of calculation
time, which is passed to the calculation of $F_2$, as well, through the mapping between $F_2-C$.
The time reduction becomes more notable as the dimension of the embedding space moves from one
to three dimensions. It is, also, negligible in the absence of error calculation, since in that
case no further reading of data is required.
The technique remains advantageous, as long as the needed number of estimation 
points remains higher than the average number of pairs per event.
 
The extreme time reduction enables us to perform scanning of data at minuscule scales $R$ or
very high partitions $M$, thus, extracting every piece of information up to experimental accuracy.

We are confident that the techniques developed here can become an indispensable tool in analysing
data which contain a very weak signal hidden in a high amount of noise.
Also, they can be useful to analyses at embedding spaces of higher dimensions.
Especially, we urge for their use in the study of correlation phenomena in heavy-ion experiments
whenever modest multiplicities per event are achieved. The analyses with the ring technique can be
repeated to all the cases where other techniques have not achieved in reaching the 
territory of either detector resolution or empty bins. 

\vspace{0.5cm}
{\bf Acknowledgement.} We wish to thank N.~Davis for fruitful discussions.

\vspace{0.5cm}
\appendix

{\bf \Large Appendix}

\numberwithin{equation}{section}

\section{Distributions for different Multiplicities} \label{sec:dif_mul}

Often the analysed data contain events with different multiplicities, $N_m$,
per event. Let $C_m$ and $F_{2,m}$ be the correlation integral and the
factorial moments of the 2nd order, respectively, for the subset of our
data with constant $N_m$. Also, let $C_t$ and $F_{2,t}$ be the correlation
integral and the
factorial moments of the 2nd order, respectively, for the whole set of our
data which contain events with different $N_m$.
We shall see how the above are related.

The fraction of the events with constant $N_m$, $N_{e,m}$, to the total
number of events, $N_e$, is
\begin{equation}
f_m = \frac{N_{e,m}}{N_e}
\end{equation}
and apparently
\begin{equation}
\sum\limits_m {f_m}  = 1\;.
\end{equation}

The average number of pairs with respect to the events which we will find at a certain
partition $M$ if we analyse all our data is
\[
\left\langle N_p(M) \right\rangle _e = \frac{1}{N_e}\sum\limits_e N_p(M)  =
\frac{1}{N_e}\sum\limits_m \sum\limits_{e_m} N_{p,m}(M) =
\]
\begin{equation}
\frac{1}{N_e}\sum\limits_m N_{e,m} \left\langle N_p(M) \right\rangle_{e,m} = 
\sum\limits_m f_m \left\langle N_p(M) \right\rangle _{e,m} \;,
\end{equation}
where with $e_m$ we denote the events with certain multiplicity $N_m$ and with
$\left\langle N_p(M) \right\rangle_{e,m}$ the average number of pairs at a certain
partition with respect to only these events. 

Likewise, the average number of pairs with respect to the events we will find at a certain
scale $R$ if we analyse all our data is
\begin{equation}
\left\langle N_p(R) \right\rangle _e =
\sum\limits_m f_m \left\langle N_p(R) \right\rangle _{e,m}\;.
\end{equation}

The correlation integral for the whole set of our data is
\[
C_t (R) = 
\frac{2 \left\langle N_p(R) \right\rangle_e}{\left\langle N_m(N_m - 1) \right\rangle_e} = 
\frac{\sum\limits_m f_m 2 \left\langle N_p(R) \right\rangle _{e,m}}{\left\langle N_m(N_m - 1) \right\rangle_e} =
\sum\limits_m f_m \frac{N_m(N_m - 1)}{\left\langle N_m (N_m - 1) \right\rangle_e}
\frac{2 \left\langle N_p(R) \right\rangle_{e,m}}{N_m(N_m - 1)}\Rightarrow
\]
\begin{equation}\label{eq:Ct-Cm}
C_t (R) = \sum\limits_m g_m C_m (R)\;\;\;,\;\;\;
g_m = f_m \frac{N_m(N_m - 1)}{\left\langle N_m (N_m - 1) \right\rangle_e}\;.
\end{equation}

The factorial moments of 2nd order for the whole set of our data is
\[
F_{2,t} (M) = 
\frac{2 \left\langle N_p(M) \right\rangle_e}{\left\langle N_m \right\rangle^2_e} = 
\frac{\sum\limits_m f_m 2 \left\langle N_p(M) \right\rangle _{e,m}}{\left\langle N_m\right\rangle^2_e} = 
\sum\limits_m f_m \frac{N_m^2}{\left\langle N_m \right\rangle^2_e}
\frac{2 \left\langle N_p(M) \right\rangle_{e,m}}{N_m^2}\Rightarrow
\]
\begin{equation}\label{eq:F2t-F2m}
F_{2,t} (M) = \sum\limits_m h_m F_{2,m} (M)\;\;\;,\;\;\;
h_m = f_m \frac{N_m^2}{\left\langle N_m \right\rangle^2_e}\;.
\end{equation}

Eqs.~(\ref{eq:Ct-Cm}) and (\ref{eq:F2t-F2m}) show the relation between the distributions
formed for subsets of our data with constant multiplicity to the whole data distribution.
We note that, while
\begin{equation}
\sum\limits_m g_m  = 
\sum\limits_m f_m\frac{N_m(N_m - 1)}{ \left\langle N_m(N_m - 1) \right\rangle_e} =
\frac{\left\langle N_m (N_m - 1) \right\rangle_e}{\left\langle N_m(N_m - 1) \right\rangle_e} = 1\;,
\end{equation}
\begin{equation}
\sum\limits_m h_m  = 
\sum\limits_m f_m\frac{N_m^2}{ \left\langle N_m \right\rangle^2_e} =
\frac{\left\langle N_m^2 \right\rangle_e}{\left\langle N_m \right\rangle^2_e} \geq 1\;.
\end{equation}

Now the fixed multiplicities distributions $C_m$ and $F_{2,m}$ are mapped
between each other according to eqs.~(\ref{eq:C-F2}) and (\ref{eq:F2-C}), thus
\begin{equation}
C_m(R) = \frac{N_m}{N_m - 1} M^{-d} F_{2,m}(M)\;.
\end{equation}

Then for the whole data distributions we have
\[
C_t(R) = \sum\limits_m g_m C_m(R) = 
\sum\limits_m f_m\frac{N_m(N_m - 1)} {\left\langle N_m(N_m - 1) \right\rangle_e} 
\frac{N_m}{N_m - 1} M^{-d} F_{2,m}(M)= 
\]
\[
=\frac{M^{-d}}{\left\langle N_m(N_m-1) \right\rangle_e}
\sum\limits_m {f_m}N_m^2\frac{2 \left\langle N_p(M) \right\rangle_{e,m}}{N_m^2} = 
\frac{M^{-d}}{\left\langle N_m(N_m-1) \right\rangle_e} 2 \left\langle N_p(M) \right\rangle_e =
\]
\begin{equation}
\frac{\left\langle N_m \right\rangle_e^2}{\left\langle N_m(N_m - 1) \right\rangle_e}
{M^{-d}}\frac{2 \left\langle N_p(M) \right\rangle_e}{\left\langle N_m \right\rangle_e^2} \Rightarrow 
C_t(R) = \frac{\left\langle N_m \right\rangle_e^2}{\left\langle N_m(N_m - 1) \right\rangle_e} M^{-d} F_{2,t}(M)\;.
\end{equation}
So, as expected, the mapping of eqs.~(\ref{eq:C-F2}) and (\ref{eq:F2-C}) can be used either for the
constant or for the varying multiplicity distributions.

We, also, consider how time consumption changes
when the multiplicity $N_m$ per event varies.
We can divide our $N_e$ in classes which
contain $N_{e,m}$ events with fixed $N_m$. Each such
class takes time $t_m$ to be analysed. 
This time, according to findings of section
\ref{sec:ring} for fixed multiplicities will be
\begin{equation}
t_m = T_0 N_{e,m} \tilde{f}_D (N_D) \tilde{f}_N (N_m)\;,
\end{equation}
where $T_0$ is a constant, $\tilde{f}_D$ is a
function of the estimation points and $\tilde{f}_N$ 
is a function of the multiplicity.
Then the time $t$ to analyse the whole set will be
\[
t=\sum\limits_m t_m = \sum\limits_m T_0 N_{e,m} \tilde{f}_D (N_D) \tilde{f}_N (N_m)=
T_0 \tilde{f}_D (N_D) \sum\limits_m N_{e,m}  \tilde{f}_N (N_m)=
\]
\begin{equation}
=T_0 N_e \tilde{f}_D (N_D) \sum\limits_m \frac{N_{e,m}}{N_e}  \tilde{f}_N (N_m)
=T_0 N_e \tilde{f}_D (N_D) \sum\limits_m f_m  \tilde{f}_N (N_m)=
T_0 N_e \tilde{f}_D (N_D) 
\left\langle \tilde{f}_N (N_m) \right\rangle_e
\end{equation}
So, the time needed to process events with 
varying multiplicities depends on the average
value, with respect to the events, of the function
of $N_m$ which corresponds to the time 
consumption for fixed $N_m$ cases.


\begin{thebibliography}{}

\bibitem{mom1}
A.~Bialas, R.~Peshanski, Nucl.~Phys.~B {\bf 273} (1986) 703.

\bibitem{mom2}
A.~Bialas, R.~Peshanski, Nucl.~Phys.~B {\bf 308} (1988) 857.

\bibitem{inter1}
J.~Wosiek, Acta Phys.~Polon.~B {\bf 19} (1988) 863.

\bibitem{inter2}
H.~Satz, Nucl.~Phys.~B {\bf 326} (1989) 613.

\bibitem{inter3}
N.G.~Antoniou, Phys.~Lett.~B {\bf 245} (1990) 624.

\bibitem{inter4}
Ph.~Brax, R.~Peschanski, Phys.~Lett.~B {\bf 346} (1990) 65.

\bibitem{inter5}
A.~Bialas, R.C.~Hwa, Phys.~Lett.~B {\bf 253} (1991) 436.

\bibitem{inter6}
M.~Ploszajczak, A.~Tucholski, P.~Bozek, Phys.~Lett.~B {\bf 262} (1991) 383.

\bibitem{inter7}
R.C.~Hwa, M.T.~Nazirov, Phys.~Rev.~Lett.~{\bf 69} (1992) 741.

\bibitem{inter8}
I.M.~Dremin, M.T.~Nazirov, Z.~Phys.~C {\bf 59} (1993) 647.

\bibitem{inter9}
E.A.~De Wolf, I.M.~Dremin, W.~Kittel, Phys.~Rep.~{\bf 270} (1996) 1.

\bibitem{inter10}
X.~Cai, C.B.~Yang, Z.M.~Zhou, Phys.~Rev.~C {\bf 54} (1996) 2775.

\bibitem{inter11}
N.G.~Antoniou, F.K.~Diakonos, C.N.~Ktorides, M.~Lahanas, Phys.~Lett.~B {\bf 432} (1998) 8.

\bibitem{inter12}
N.G.~Antoniou, Y.F.~Contoyiannis, F.K.~Diakonos, A.I.~Karanikas,
C.N.~Ktorides, Nucl.~Phys.~A {\bf 693} (2001) 799.

\bibitem{inter13}
N.G.~Antoniou, F.K.~Diakonos, E.~Saridakis, Phys.~Rev.~C {\bf 78} (2008) 024908.

\bibitem{inter14}
R.C.~Hwa, C.B.~Yang, Phys.~Rev.~C {\bf 85} (2012) 044914.

\bibitem{inter15}
X.Z.~Bai, C.B.~Yang, Int.~J.~Mod.~Phys.~E {\bf 22} (2013) 1350059.

\bibitem{inter16}
R.C.~Hwa, C.B.~Yang, Acta Phys.~Pol.~B {\bf 48} (2016) 23.

\bibitem{Crit-Opal}
N.G.~Antoniou, F.K.~Diakonos, A.S.~Kapoyannis, K.S.~Kousouris, Phys.~Rev.~Lett.~{\bf 97} (2006) 032002.

\bibitem{CI} 
P.~Grassberger, I.~Procaccia, Physica D {\bf 9} (1983) 189.

\bibitem{NA49_protons}
T.~Anticic {\it et al.}, Eur.~Phys.~J.~C {\bf 75} (2015) 587.

\bibitem{Antoniou_PRC_2016}
N.G.~Antoniou, N.~Davis, F.~K.~Diakonos, Phys.~Rev.~C {\bf 93} (2016) 014908.

\bibitem{Antoniou_jopg_2019} N.G.~Antoniou, F.K.~Diakonos, J.~Phys.~G: Nucl.~Part.~Phys.~{\bf 46} (2019) 035101.

\bibitem{Levy}
P.A.~Alemany, D.H.~Zanette, Phys.~Rev.~E {\bf 49} (1994) R956.

\bibitem{Hen}
M.~Hénon, Commun.~in Math.~Phys.~{\bf 50} (1976) 69. 

\bibitem{Ike}
K.~Ikeda, Optics Commun.~{\bf 30} (1979) 257.

\bibitem{Lor1}
E.N.~Lorenz, J.~Atmos.~Sci.~{\bf 20} (1963) 130.

\bibitem{Lor2}
D.~Viswanath, Physica D {\bf 190} (2004) 115.

\bibitem{dimU1-H-L}
J.C.~Sprott, G.~Rowlands, IJBC, {\bf 11} (2001) 1865.

\bibitem{dimH-L}
P.~Grassberger, I.~Procaccia, Phys.~Rev.~Lett.~{\bf 50} (1983) 346.

\bibitem{dimH-I-L}
J.~Jaquette, B.~Schweinhart, Commun.~Nonlinear Sci.~Numer.~Simul.~{\bf 84}
(2020) 105163, [arXiv:1907.11182].

\end{thebibliography}
\end{document}